\documentclass[aps,showpacs,twocolumn,twoside,pre,10pt]{revtex4-1}
\usepackage{amsmath} 
\usepackage{amssymb}
\usepackage{graphicx}
\usepackage{color}      
\usepackage{hyperref}   
\def\vecb#1{\boldsymbol{#1}}
\def\ket#1{|#1\rangle}
\def\bra#1{\langle#1|}
\def\scal#1#2{\langle#1|#2\rangle}
\def\matr#1#2#3{\langle#1|#2|#3\rangle}
\def\abs#1{\left\lvert#1\right\rvert}

\def\ave#1{\langle#1\rangle}

\def\dis#1{\langle\langle#1^2\rangle\rangle}
\def\={\!=\!}
\def\>{\!>\!}
\def\<{\!<\!}
\def\-{\!-\!}
\def\+{\!+\!}
\def\tto{\!\to\!}
\def\abs#1{\left|#1\right|}

\def\uvo#1{\lq\lq #1\rq\rq}
\def\Eqref#1{Eq.\,\eqref{#1}}
\def\Eqsref#1{Eqs.\,\eqref{#1}}
\def\Figref#1{Fig.\,\ref{#1}}
\def\Figsref#1{Figs.\,\ref{#1}}

\def\QPT{\textsc{qpt}}
\def\QPTs{\textsc{qpt}s}

\def\EP{\textsc{ep}}
\def\EPs{\textsc{ep}s}
\def\DP{\textsc{dp}}
\def\DPs{\textsc{dp}s}
\def\GOE{\textsc{goe}}
\def\GUE{\textsc{gue}}

\begin{document}

\title{Exceptional points near first- and second-order quantum phase transitions}
\author{Pavel Str{\'a}nsk{\'y}, Martin Dvo{\v r}{\'a}k, Pavel Cejnar}
\affiliation{Institute of Particle and Nuclear Physics, Faculty of Mathematics and Physics, Charles University, 
  V~Hole{\v s}ovi{\v c}k{\' a}ch 2, 180\,00 Prague, Czechia}
\date{\today}

\begin{abstract}
We study impact of quantum phase transitions (\QPTs) on the distribution of exceptional points (\EPs) of the Hamiltonian in complex-extended parameter domain.
Analyzing first- and second-order \QPTs\ in the Lipkin model, we find an exponentially and polynomially close approach of \EPs\ to the respective critical point with an increasing size of the system.
If the critical Hamiltonian is subject to random perturbations of various kinds, the averaged distribution of \EPs\ close to the critical point still carries decisive information on the \QPT\ type.
We therefore claim that properties of the \EP\ distribution represent a parametrization-independent signature of criticality in quantum systems.
\end{abstract}

\maketitle

\section{Introduction}

Almost all quantum mechanical problems depend on some parameters---external field strengths, internal coupling constants etc.
Various choices of these parameters may lead to dramatically different solutions.
In some systems, the variation of solutions with parameters may even have a critical character, which means that in the infinite-size limit it becomes abrupt, nonanalytic at some particular parameter values.
We encounter various types of ground-state or excited-state {\em quantum phase transitions\/} (\QPTs) \cite{Sac11,Car11,Cej16}. 
Do we understand the internal mechanisms behind this kind of behavior?
Can we predict in which parameter domains it can be expected?

In particular, consider a Hamiltonian $\hat{H}\equiv\hat{H}({\lambda})$ depending linearly on a single real control parameter $\lambda$:
\begin{equation}
\hat{H}(\lambda)=\hat{H}(0)+\lambda\hat{V}
\,.
\label{Hlin}
\end{equation}
Here, $\hat{H}(0)$ is a \uvo{free} Hamiltonian and $\hat{V}$ an arbitrary \uvo{interaction}, both associated with Hermitian, in general noncommuting operators represented by real matrices of a finite dimension $d$. 
Elementary analysis reveals that abrupt variations of eigenfunctions of an arbitrary operator take place at its degeneracy points where two (or eventually more) eigenvalues join.
Assuming $\hat{H}({\lambda})$  with no hidden symmetry (i.e., acting  irreducibly in the whole Hilbert space or taken in a single irreducible subspace), we know that almost all crossings of energy levels $E_n(\lambda)$ should be avoided \cite{Neu29}.
However, the true degeneracy points $E_n\=E_{n'}$ can be found  in the plane of {\em complex\/} $\vecb{\lambda}\equiv\lambda\+i\mu$, that is, for a non-Hermitian extension of the Hamiltonian \cite{Moi11}.

The non-Hermitian degeneracies, so-called {\em exceptional points\/} (\EPs) \cite{Kat66}, have a different character than ordinary degeneracies of Hermitian operators.
While an ordinary degeneracy (so-called diabolic point, \DP) in a 2-dimensional (or more) parameter space is just a conical intersection of two Hamiltonian eigenvalues \cite{Ber84}, a generic \EP\ represents the square-root type of branch point connecting two Riemann sheets of the eigenvalue solution in the plane $\vecb{\lambda}\in\mathbb{C}$ \cite{Ben69,Moi80,Zir83,Hei90,Hei91}.
Since any pair of real energies can be continuously linked up by an appropriate loopy path encircling various \EPs\ in the complex plane, the whole energy spectrum becomes a single entangled object allowing no strict distinction between different levels.
The eigenvectors at the degeneracy points $\vecb{\lambda}^{\rm ep}_i$ do not form a complete basis and the single eigenvector associated with the pair of coalescing levels becomes selforthogonal \cite{Moi11}.
In spite of these unusual properties (see Appendix~\ref{SeApA}), the locations of \EPs\ determine the main features of the real energy spectrum and its evolution with $\lambda\in\mathbb{R}$.
In particular, the presence of $\vecb{\lambda}^{\rm ep}_i$ near the real axis shows up as a sharp avoided crossing of the corresponding levels at $\lambda\approx{\rm Re}\vecb{\lambda}^{\rm ep}_i$ and therefore induces a rapid evolution of the associated eigenstates $\ket{\psi_n(\lambda)}$.

In view of this background it is not surprising that \EPs\ play an essential role in the description of \QPTs\ \cite{Hei88,Hei02,Cej05,Cej07,Cej09,Lee14,Bor15,Sin17}.
Here we focus solely on the ground-state \QPTs, which are associated with sudden changes of the  ground-state energy $E^{\rm gs}\=\matr{\psi^{\rm gs}}{\hat{H}}{\psi^{\rm gs}}$ (where $\ket{\psi^{\rm gs}}$ stands for the ground-state wave function) and order parameter $\ave{O}^{\rm gs}\=\matr{\psi^{\rm gs}}{\hat{O}}{\psi^{\rm gs}}$ (with $\hat{O}$ standing for an operator associated with a suitable observable characterizing the ground-state structure)  in a vicinity of a certain critical Hamiltonian $\hat{H}^{\rm c}$, for instance at a particular value $\lambda^{\rm c}$ of the control parameter in \Eqref{Hlin}.
We shall stress that the \QPTs, similarly to thermal phase transitions, become truly nonanalytic only in the limit of the system's infinite size, $N\tto\infty$.
It turns out that as the size increases, some of the \EPs\ converge to the \QPT\ critical point $\lambda^{\rm c}$ on the real axis of $\vecb{\lambda}$, in analogy with the behavior of complex zeros of the thermodynamic partition function near thermal phase transitions \cite{Cej05,Cej07,Yan52,Bor00}.

In a first-order (discontinuous) \QPT,  the order parameter exhibits a discontinuity.
Typical examples are systems with the potential energy dependence $V(x)$ having the form of a double well.
The crossing of both potential minima at a certain $\lambda\=\lambda^{\rm c1}$, with $\hat{H}(\lambda^{\rm c1})\=\hat{H}^{\rm c1}$ describing a degenerate double-well system, indicates a jump of the global minimum from one well to the other.
The order parameter, which in this case can be the average coordinate $\ave{x}^{\rm gs}$, changes abruptly at the critical point  between the values corresponding to the momentary localizations of both minima.

On the other hand, in a continuous (second-order or more general) \QPT\ the order parameter is a continuous function of control parameter and the singularity is shifted to the first or higher derivatives.
This typically happens if the potential energy $V(x)$ develops at $\lambda\=\lambda^{\rm c2}$ a degenerate (higher than quadratic) global minimum, so the critical Hamiltonian $\hat{H}(\lambda^{\rm c2})\=\hat{H}^{\rm c2}$ exhibits an accumulation of eigenstates near the lowest energy.
An infinitesimal change of $\lambda$ lifts the degeneracy, transforming the minimum into one or more quadratic stationary points.
The order parameter $\ave{x}^{\rm gs}$ then varies in a continuous way, but with discontinuous or infinite derivatives in variable $\lambda$.

In this paper, we study the distribution of \EPs\ in $\vecb{\lambda}\in\mathbb{C}$ for Hamiltonians of the form \eqref{Hlin} near critical points $\lambda^{{\rm c}k}\in\mathbb{R}$ of generic first-order ($k\=1$) and second-order ($k\=2$) \QPTs.
We search for the features of the \EP\ distribution and its dependence on the system's size that are distinctive for the transition type. 
The plan is as follows: 
In Sec.\,\ref{SeLi}, we present results for specific families of Hamiltonians within the Lipkin model, making an example of both the above \QPT\ types.
In Sec.\,\ref{SeRa}, we investigate the \EP\ distributions associated with random perturbations $\lambda\hat{V}$ of a critical Hamiltonian $\hat{H}(0)\equiv\hat{H}^{{\rm c}k}$ taken at the first- and second-order \QPT.
We argue that the critical Hamiltonians of either type have some general characteristics (reducible to the associated \EP\ distributions) that go beyond any particular model-dependent Hamiltonian parametrization. 
Sec.\,\ref{SeCo} makes a summary of results.

\section{Exceptional points for critical Hamiltonians in the Lipkin model}
\label{SeLi}

In this Section, we illustrate the distribution of \EPs\ around a first- and second-order \QPT\ in the model of Lipkin, Meshkov and Glick \cite{Lip66}, here shortened as the \uvo{Lipkin model}.
It was originally introduced as a toy model for nuclear physics, but recent experimental results \cite{Gro10} induced renewed attention to this model in the context of cold atoms and general many-body physics.

\subsection{Hamiltonian and ground-state critical properties}

The Lipkin model can be introduced in several alternative ways. 
It was originally formulated as a system of $N$ interacting fermions on two energy levels, but it can be cast also in terms of two interacting bosonic species, or through a system of $N$ interacting spin-$\tfrac{1}{2}$ particles or two-level atoms.
Following the latter representation, we assign to each (the $l$th) spin/atom a 2-dimensional Hilbert space ${\cal H}^{(l)}$ 
and the set $\vec{\hat{\sigma}}^{(l)}\!\equiv\!(\hat{\sigma}^{(l)}_1,\hat{\sigma}^{(l)}_2,\hat{\sigma}^{(l)}_3)$ of Pauli matrices acting on it.
The collective spin operators $\hat{\vec{J}}\!\equiv\!(\hat{J}_1,\hat{J}_2,\hat{J}_3)$ on the full $2^N$-dimensional Hilbert space ${\cal H}\!\equiv\!\otimes_{l=1}^{N}{\cal H}^{(l)}$ are defined as $\hat{\vec{J}}=\sum_{l=1}^{N}\vec{\hat{\sigma}}^{(l)}$ and satisfy the usual SU(2) commutation rules.

The Lipkin Hamiltonian $\hat{H}$ is supposed to be written solely in terms of the collective spin operators $\hat{\vec{J}}$, or equivalently $\hat{J}_{\pm}\=\hat{J}_1\pm i\hat{J}_2$ and $\hat{J}_0\=\hat{J}_3$. 
It therefore conserves the $\hat{\vec{J}}^2$ quantum number $j$.
The full Hilbert space ${\cal H}$ splits into a sum of subspaces with fixed $j\=\{j^{\rm min},\dots,j^{\rm max}\}$, where $j^{\rm min}\=0$ or $\frac{1}{2}$ for $N$ even or odd, respectively, and $j^{\rm max}\=\tfrac{N}{2}$ (the value $2j$ represents a number of excitable spins).
These subspaces, except the unique one with $j\=j^{\rm max}$, appear in a large number of replicas differing by the inherent exchange symmetry of the state vectors involved (see e.g. Ref.\,\cite{Cej16}).
Since each of these $(2j\+1)$-dimensional subspaces is invariant under the action of $\hat{H}$, the dynamics can be restricted to any of them.
The usual choice, which we also follow here, is the fully exchange-symmetric subspace with $j\=j^{\rm max}$ and dimension $d\=N\+1$.

An arbitrary Lipkin Hamiltonian restricted to any of the fixed-$j$ subspaces represents a system with one degree of freedom that can be transformed to the coordinate--momentum form.
One can use, e.g., the Holstein-Primakoff mapping \cite{Hol40} of the collective spin operators:
\begin{equation}
\left(\hat{J}_-,\hat{J}_0,\hat{J}_+\right)\mapsto\left(\sqrt{2j\-\hat{b}^{\dag}\hat{b}}\ \hat{b},\ \hat{b}^{\dag}\hat{b}\-j,\ \hat{b}^{\dag}\sqrt{2j\-\hat{b}^{\dag}\hat{b}}\right)
\label{HPmap}
\end{equation}
followed by the transformation of boson creation and annihilation operators $\hat{b}^{\dag},\hat{b}$ to coordinate and momentum operators $\hat{x},\hat{p}$:
\begin{equation}
\left(\hat{b}^{\dag},b\right)\mapsto\sqrt{j}\,\bigl(\hat{x}\-i\hat{p},\hat{x}\+i\hat{p}\bigr)
\label{XPmap}\,.
\end{equation}
The commutation relation $[\hat{x},\hat{p}]\=i/2j$ indicates that the quantity $1/2j$ plays the role of an effective Planck constant.
In the limit $j\tto\infty$ (hence also $N\tto\infty$), the Hamiltonian $\hat{H}$ with substitutions \eqref{HPmap} and \eqref{XPmap} becomes a function $H$ of commuting variables $x$ and $p$ satisfying $x^2\+p^2\!\leq\!2$, which defines the classical phase space associated with the model. 

The Lipkin model with $N,j\tto\infty$ exhibits several ground-state phase transitions that show up as nonanalytic changes of the absolute minimum of function $H(x,p)$ with varying model control parameters, see e.g. Refs.\,\cite{Gil78,Cas06,Vid06,Rib08}
(and \cite{Cej16} for an outline).
To demonstrate these effects, we represent the Hamiltonian close to the respective \QPT\ in the form \eqref{Hlin}, i.e., as $\hat{H}^{{\rm qpt}k}(\lambda)$  with $k\=1,2$ and a single control parameter $\lambda$ passing through a certain critical value $\lambda^{{\rm c}k}$.

A possible Hamiltonian $\hat{H}^{\rm qpt1}(\lambda)$ with the {\em first-order\/} \QPT\ has: 
\begin{equation}
\hat{H}^{\rm qpt1}(0)\=\hat{J}_3\-\frac{a}{j}\,\hat{J}_1^2
\,,\ \
\hat{V}^{\rm qpt1}\=-J_1\-\frac{1}{2j}\left(\hat{J}_1\hat{J}_3\+\hat{J}_3\hat{J}_1\right),
\label{HL1}
\end{equation}
where $a\>\frac{1}{2}$ is a tunable constant, in the following set to $a\=3$.
There is an apparent symmetry of the spectrum of $\hat{H}^{\rm qpt1}(\lambda)$ under the inversion $\lambda\tto -\lambda$ (the corresponding Hamiltonians differ just by $\pi$-rotation around the 3rd axis).
So if $\lambda$ crosses the critical value $\lambda^{\rm c1}\=0$, the ground-state expectation value $\ave{J_1}^{\rm gs}\equiv\matr{\psi^{\rm gs}}{\hat{J}_1}{\psi^{\rm gs}}$ changes its sign.
The change gets sharper with increasing $N$ and tends to a sudden flip with $N\tto\infty$.
Indeed, writing down the classical Hamiltonian associated with $\hat{H}^{\rm qpt1}(\lambda)$:
\begin{equation}
\frac{H^{\rm qpt1}}{2j}\=\frac{1\-2a}{2}x^2\-\frac{\lambda}{2}x^3\sqrt{2\-x^2}+\frac{a}{2}x^4\+K\-\frac{1}{2}
\label{HL1cl}\,,
\end{equation}
where $K(x,p)$ is a complicated (position-dependent and quartic in momentum) kinetic term not given explicitly here, we immediately see that the classical Hamiltonian $H^{\rm c1}$ associated with the quantum critical Hamiltonian $\hat{H}^{\rm c1}\!\equiv\!\hat{H}^{\rm qpt1}(\lambda^{\rm c1})$ corresponds to a degenerate double-well system which is {\em parity symmetric\/}.
The quantity $\ave{J_1}^{\rm gs}\propto\ave{x\sqrt{2\-x^2\-p^2}}^{\rm gs}$ can be seen as an order parameter characterizing the ground-state \uvo{phases} in the present \QPT.

The Lipkin Hamiltonian with a {\em second-order\/} \QPT\ can be written as $\hat{H}^{\rm qpt2}(\lambda)$ with:
\begin{equation}
\hat{H}^{\rm qpt2}(0)=\hat{J}_3
\,,\quad
\hat{V}^{\rm qpt2}=-\frac{1}{2j}\hat{J}_1^2
\,.\label{HL2}
\end{equation}
The order parameter might be again associated with $\ave{J_1}^{\rm gs}$, but a more suitable choice is the ground-state spin inversion parameter $\ave{I}^{\rm gs}\equiv\matr{\psi^{\rm gs}}{\hat{J}_3\+j}{\psi^{\rm gs}}\propto\ave{x^2\+p^2}^{\rm gs}$.
For $\lambda$ below the value $\lambda^{\rm c2}\=1$ we obtain $\ave{I}^{\rm gs}\=0$, which means that all spins point down in the lowest state, while above $\lambda^{\rm c2}$ we find a nonzero (increasing with $\lambda$) value $\ave{I}^{\rm gs}$, indicating a measurable fraction of spin-up orientations.
The change of $\ave{I}^{\rm gs}$ is continuous, but for $N\tto\infty$ the first derivative $\frac{d}{d\lambda}\ave{I}^{\rm gs}$ varies discontinuously at $\lambda^{\rm c2}$.
The classical Hamiltonian
\begin{equation}
\frac{H^{\rm qpt2}}{2j}=\frac{1\-\lambda}{2}x^2+\frac{\lambda}{4}x^4+\frac{2\+\lambda x^2}{4}p^2-\frac{1}{2}
\label{HL2cl}
\end{equation}
corresponding to \Eqref{HL2} shows that the critical Hamiltonian $H^{\rm c2}$ associated with $\hat{H}^{\rm c2}\!\equiv\!\hat{H}^{\rm qpt2}(\lambda^{\rm c2})$ is a pure quartic oscillator with a position-dependent kinetic term.

The form \eqref{HL2} gives us yet another possibility to create a first-order \QPT.
The corresponding Hamiltonian $\hat{H}^{\rm qpt1'}(\lambda)$ is determined by:
\begin{equation}
\hat{H}^{\rm qpt1'}(0)=\hat{J}_3\,,\quad
\hat{V}^{\rm qpt1'}=-\frac{1}{2j}\left[\hat{J}_1\+c\left(\hat{J}_3\+j\right)\right]^2
\label{HL1'},
\end{equation}
where the interaction term is modified with respect to \Eqref{HL2} and brings a new parameter $c$ (in the following fixed at $c\=4$).
The order parameter characterizing the relevant phases is again the ground-state spin inversion $\ave{I}^{\rm gs}$, which for $c\!\neq\!0$ changes from zero to a nonzero value in an abrupt, discontinuous way at $\lambda\=\lambda^{\rm c1'}\=1/(1\+c^2)$.
The classical Hamiltonian corresponding to \Eqref{HL1'} is:
\begin{equation}
\frac{H^{\rm qpt1'}}{2j}\=\frac{1\-\lambda}{2}x^2-\frac{c\lambda}{2}x^3\sqrt{2\-x^2}+\frac{\lambda(1\-c^2)}{4}x^4+K'\-\frac{1}{2}
\label{HL1'cl},
\end{equation}
where $K'(x,p)$ is again a certain kinetic term.
We see that the classical critical Hamiltonian $H^{\rm c1'}$ associated with $\hat{H}^{\rm c1'}\!\equiv\!\hat{H}^{\rm qpt1'}(\lambda^{\rm c1'})$ corresponds again to a degenerate double-well system, but now {\em parity asymmetric}, in contrast to the previous first-order \QPT\ case $\hat{H}^{\rm c1}$.
 
\subsection{Distributions of exceptional points}

Prior to discussing the \EP\ distributions associated with the above critical Hamiltonians, we have to comment on the general methods for finding the \EPs.
A straightforward way is to search roots of a polynomial $D(\vecb{\lambda})$ obtained by elimination of the system of equations
\begin{equation}
{\rm det}[\hat{H}(\vecb{\lambda})\-E]=0\,,\quad
\frac{\partial}{\partial E}{\rm det}[\hat{H}(\vecb{\lambda})\-E]=0
\label{dets}
\end{equation} 
where the first equation is the eigenvalue condition and the second the degeneracy condition \cite{Zir83,Hei90,Hei91}.
Since the order of $D(\vecb{\lambda})$ is $d(d\-1)$ and its coefficients are real for Hamiltonians of the form \eqref{Hlin}, the \EPs\ come as complex conjugate pairs $(\vecb{\lambda}_i^{\rm ep},{\vecb{\lambda}_i^{\rm ep}}^*)$ with $i\=1,...,{\cal I}$, where ${\cal I}\=d(d\-1)/2$.
However, this method requires an extremely high evaluation precision and works (with commonly available computational platforms) only for moderate dimensions, say $d\!\lesssim\!30$ \cite{Dvo15}.

\begin{figure}[!t]
\includegraphics[width=\linewidth]{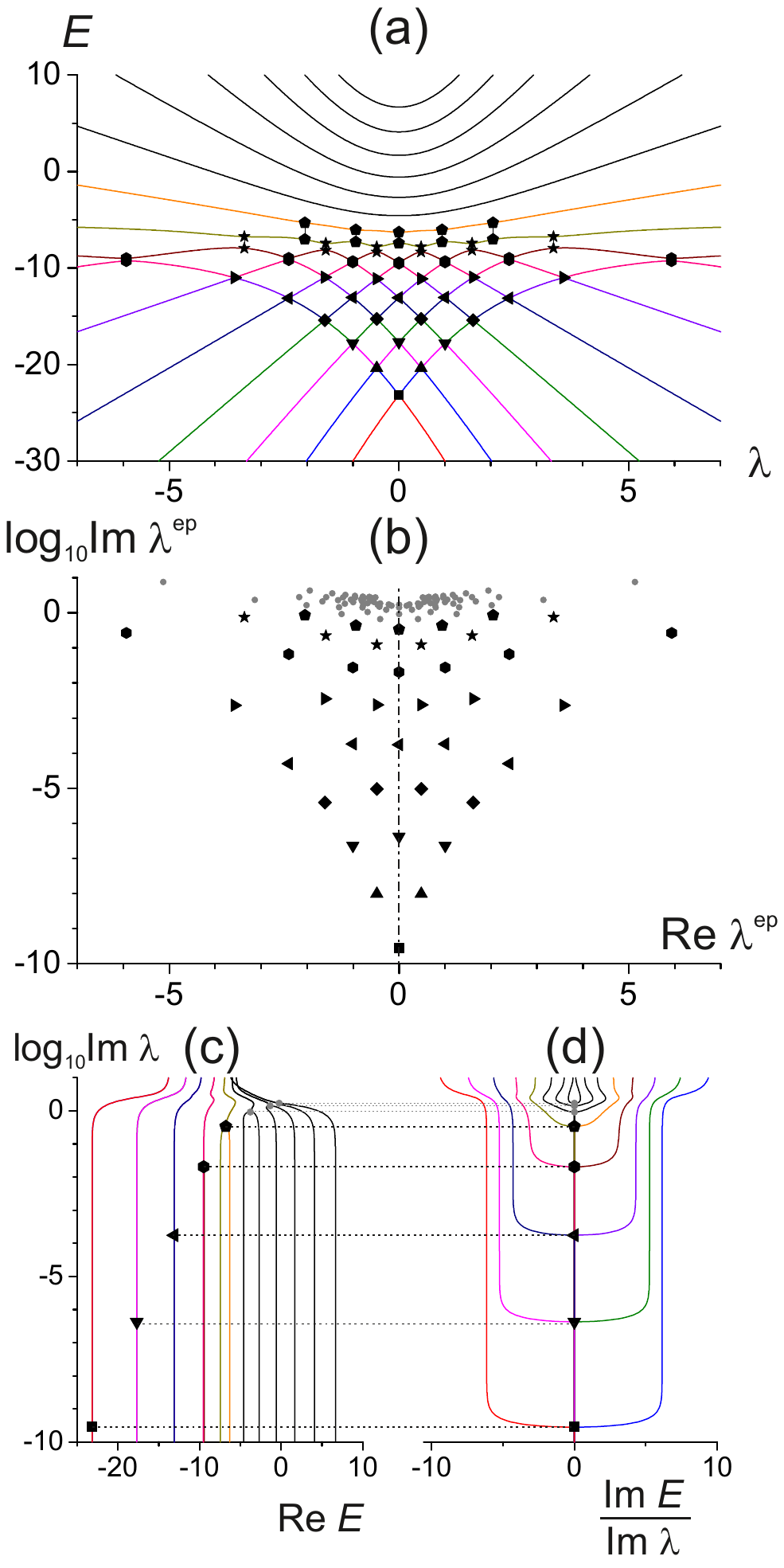}
\caption{
Energy spectrum and \EPs\ for the first-order \QPT\  Lipkin Hamiltonian \eqref{HL1} with $N\!=\!15$.
Panel (a): The $\lambda\!\leftrightarrow\!-\lambda$ symmetric energy spectrum with the avoided crossings of levels demarcated by dots.
Panel (b): The \EP\ pattern, in which selected \EPs\ are assigned to the corresponding avoided crossing in panel (a) by the dot types.
Panels (c) and (d): The evolution of real and imaginary parts of individual level energies $E_n(\vecb{\lambda})$ along a path $\vecb{\lambda}\!=\!0\+i\mu$. 
The real parts merge (panel c) and the imaginary parts diverge (panel d) as the path crosses individual \EPs\ in panel (b) [mergers in panel (c) are invisible due to very small energy differences, while in panel (d) they are emphasized by $1/{\rm Im}\lambda$ scaling of ${\rm Im}E$].
}
\label{FL1}
\end{figure}

More efficient methods have been proposed, see e.g. Ref.\,\cite{Uzd10} and the references therein, but they aim mostly at finding a single \EP\ inside a limited parameter domain.
In contrast, our task is to find all \EPs\ in a large region of $\vecb{\lambda}$.
To this end, we use a modification of the loop-integration method proposed in Ref.\,\cite{Zir83}.
The method makes use of the fact that two complex energies $E_{n}(\vecb{\lambda})$ and $E_{n'}(\vecb{\lambda})$ at a small distance $\vecb{\delta}\=\vecb{\lambda}\-\vecb{\lambda}^{\rm ep}_i$ from their associated \EP\ behave as $E_n\-E_{n'}\propto\sqrt{\vecb{\delta}}$, see Appendix~\ref{SeApA}.
Therefore, following a closed loop around the \EP, the energies $E_n$ and $ E_{n'}$ swap.
Note that here we do not take into account rare but possible cases of multiple \EPs\ connecting three or more levels \cite{Dem12}.
Generalizing the above conclusion to regions with an arbitrary number $L$ of ordinary \EPs, we observe that after closing a loop around this region, $L$ energies in the set $\{E_1,...,E_d\}$ must swap.
This makes it possible to detect large clusters of \EPs\ and by reducing the loop sizes (while keeping a sufficient precision of movements along the loops) to iteratively localize individual \EPs\ inside these clusters.

The distribution of \EPs\ for the Lipkin model was previously calculated for the second-order \QPT\ Hamiltonian similar to that in \Eqref{HL2} \cite{Sin17,Hei05}.
On the other hand, the first-order \QPT\ Hamiltonians \eqref{HL1} and \eqref{HL2} were not studied.
We start with the symmetric case $\hat{H}^{\rm qpt1}$ from \Eqref{HL1}.
The corresponding energy spectrum and a pattern of \EPs\ are depicted in \Figref{FL1}.
As explained above, the distribution of \EPs\ is symmetric under the complex conjugation, so we always show only the ${\rm Im}\vecb{\lambda}\>0$ halfplane.
The additional symmetry of the pattern in panel (b) under the real axis inversion results from the $\lambda\!\leftrightarrow\!-\lambda$ symmetry of the Hamiltonian.
We note that the imaginary axis of $\vecb{\lambda}$ in panel (b) is logarithmic, so the distances of the closest \EPs\ to the real axis are indeed very small and differ between each other by several orders of magnitude.
As discussed below, this is very typical for the first-order \QPTs.

\begin{figure}[!t]
\includegraphics[width=\linewidth]{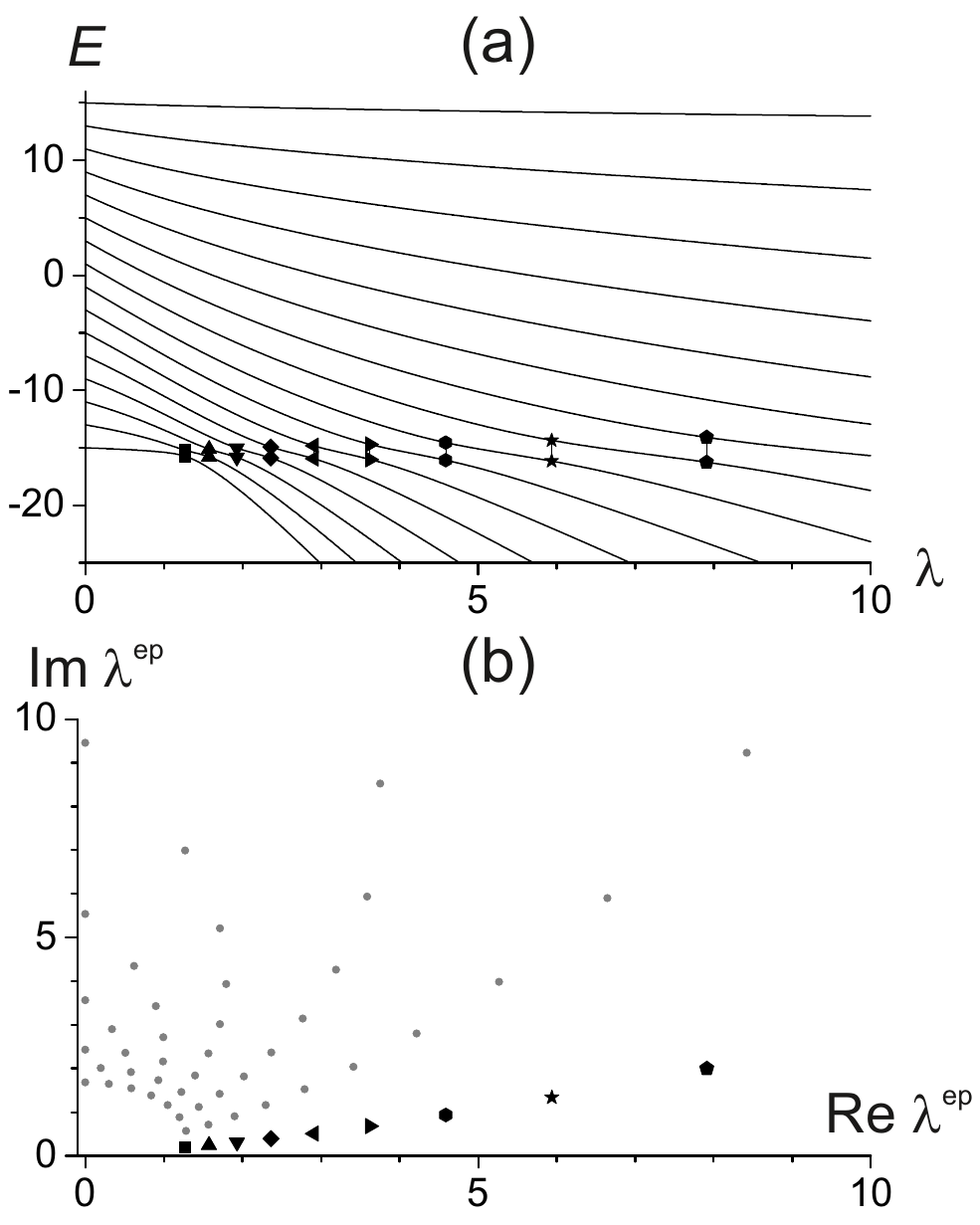}
\caption{
Energy spectrum (panel a) and \EPs\ (panel b) for the second-order \QPT\  Lipkin Hamiltonian \eqref{HL2} with $N\!=\!15$.
Only the \EPs\ closest to the real axis are assigned to the corresponding avoided crossings. 
}
\label{FL2}
\end{figure}

A comparison of the \EP\ pattern in panel (b) of \Figref{FL1} with the spectrum in panel (a) demonstrates a one-to-one correspondence of a large subset of \EPs\ with avoided crossings of real energy levels.
This is visualized by using the same dot type for the \EP\ and its associated avoided crossing. 
The assignment can be done by tracing the evolution of energies $E_n(\vecb{\lambda})$ from the avoided crossing on ${\rm Im}\vecb{\lambda}\=0$ along a straight path perpendicular to the real axis.
In panels (c) and (d) we select the line starting at $\lambda\=0$, where the spectrum shows several avoided crossings.
As ${\rm Im}\vecb{\lambda}$ increases and the path crosses locations of individual \EPs, we observe that real parts of selected energies merge and imaginary parts diverge.
This indicates a connection of the given pair of levels with the particular \EP.
As shown in Appendix~\ref{SeApA}, for an isolated pair of \EPs\  located at $\vecb{\lambda}^{\rm ep}_i$ and ${\vecb{\lambda}^{\rm ep}_i}^*$ not far from the real axis, the real energies $E_{n}$ and $E_{n'}$ corresponding to levels $n$ and $n'$ associated with the \EP\ satisfy the relation:
\begin{equation}
E_{n}(\lambda)\-E_{n'}(\lambda)=2{\cal F}_{nn'}(\lambda)\abs{\lambda\-\vecb{\lambda}^{\rm ep}_i}
\label{Edif}\,,
\end{equation}
where ${\cal F}_{nn'}(\lambda)$ is a certain regular function.
This relation holds for $\abs{\lambda\-\vecb{\lambda}^{\rm ep}_i}$ less than the radius of convergence $R$ of the Puiseux expansion (distance of the given \EP\ to the closest \EP\ involving any of levels $n,n'$, see Appendix~\ref{SeApA}), 
that is within an interval $\abs{\lambda\-{\rm Re}\vecb{\lambda}^{\rm ep}_i}\<(R^2\-{\rm Im}^2\vecb{\lambda}^{\rm ep}_i)^{1/2}$.
Assuming that ${\cal F}_{nn'}$ varies slowly on this interval, we see that the minimal spacing between the two levels  is reached at $\lambda\approx{\rm Re}\vecb{\lambda}^{\rm ep}_i$ and takes a value $\abs{E_{n}\-E_{n'}}\approx 2{\cal F}_{nn'}({\rm Re}\vecb{\lambda}^{\rm ep}_i)\abs{{\rm Im}\vecb{\lambda}^{\rm ep}_i}$ proportional to the imaginary coordinate of the \EP\ .
Indeed, a highly magnified view of the spectrum in \Figref{FL1}(a) would show that the sharpness of avoided crossings changes proportionally to the distance of the corresponding \EPs\ from the real axis.

However, an unambiguous link between the \EPs\ and avoided crossings of individual levels, as outlined above,  holds only to a limited extent.
As the \EPs\ represent square-root branch points in the system of $d$ interconnected Riemann sheets of the complex function $E(\vecb{\lambda})$, the assignment of a given \EP\ to a certain pair of real energy levels is not unique.
More precisely, it can be done only if ${\rm Im}\vecb{\lambda}^{\rm ep}_i\<R$ (the radius of convergence of the Puiseux expansion), otherwise it depends on the path we choose between the real axis and the selected \EP.
There is a large number of \EPs\ in \Figref{FL1}(b) (those demarcated by smaller, gray dots) whose assignment to the real energy levels via the path perpendicular to the real axis would not correspond to any visible avoided crossing.
The effect of these \EPs\ on the real spectrum is apparently washed out by the presence of \EPs\ with smaller values of ${\rm Im}\vecb{\lambda}^{\rm ep}_i$.
In this sense, we speak about a \uvo{screening} phenomenon.

\begin{figure}[!t]
\includegraphics[width=\linewidth]{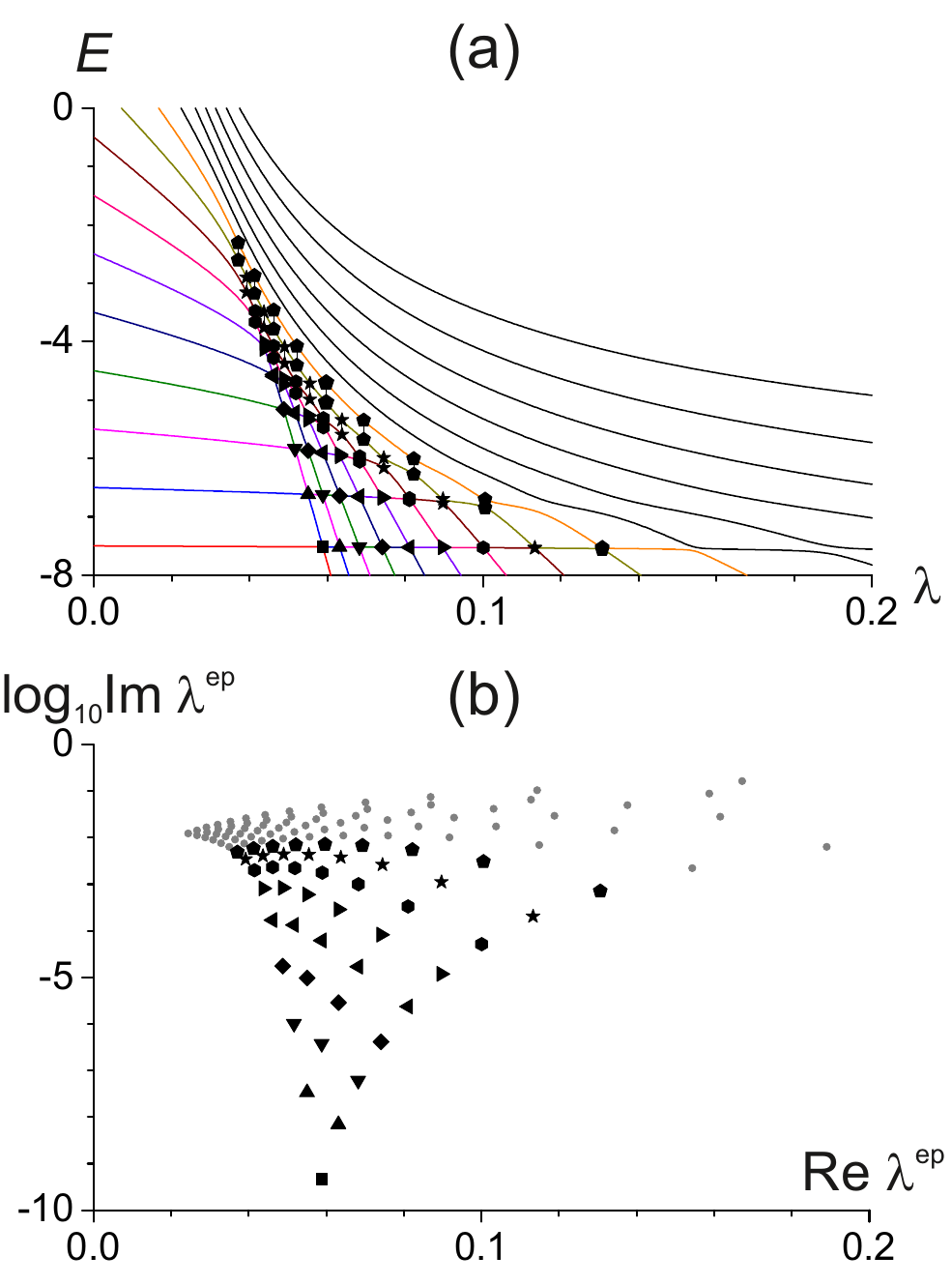}
\caption{
Energy spectrum (panel a) and \EPs\ (panel b) for the first-order \QPT\  Lipkin Hamiltonian \eqref{HL1'} with $N\!=\!15$.
}
\label{FL1'}
\end{figure}

Figure~\ref{FL2} displays the energy spectrum and a distribution of \EPs\ for the second-order \QPT\ Hamiltonian \eqref{HL2}.
The assignment of \EPs\ to real avoided crossings is now performed only for the first row of \EPs\ close to the real axis.
The pattern of \EPs\ in panel (b) is well known from Refs.\,\cite{Sin17,Hei05}.
Note that if presented also for ${\rm Re}\lambda\<0$, the pattern would be mirror symmetric with respect to ${\rm Re}\lambda\=0$; this is due to an \uvo{accidental} unitary relation between $\hat{H}^{\rm qpt2}(+\lambda)$ and $-\hat{H}^{\rm qpt2}(-\lambda)$. 
We stress that the imaginary axis of $\vecb{\lambda}$ in panel (b) of Fig.\,\ref{FL2}, in contrast to Fig.\,\ref{FL1},  is linear. 
This indicates much larger distances of \EPs\ from the real axis for the second-order \QPT\ in comparison with the first-order \QPT, and simultaneously much smaller relative differences in these distances between individual \EPs.
Based on \Eqref{Edif}, analogous statements can be formulated for spacings between individual real energy levels undergoing avoided crossings near the \QPT\ critical point.
These features are not restricted just to the present particular cases, but constitute a general distinction between the two \QPT\ types.

The last sentence is supported by Fig.\,\ref{FL1'}, which depicts the energy spectrum and the pattern of \EPs\ for the parity-asymmetric version of the first-order \QPT\ Hamiltonian, see \Eqref{HL1'}.
The main features of the \EP\ distribution, in particular a very close approach of the nearest \EPs\ to the real axis, are qualitatively similar to the previous first-order \QPT\ case in Fig.\,\ref{FL1}.
Note however that in contrast to the previous case, the present \EP\ distribution lacks the mirror symmetry around the ${\rm Re}\lambda\=\lambda^{\rm c1'}\!\doteq\!0.059$ line (there is no unitary relation between the $\lambda\=\lambda^{\rm c1'}\!\pm\delta$ Hamiltonians) and the exact centering of some \EPs\ at the critical point (the double-well system at $\lambda\=\lambda^{\rm c1'}$ is degenerate but not symmetric).

\begin{figure}[!t]
\includegraphics[width=\linewidth]{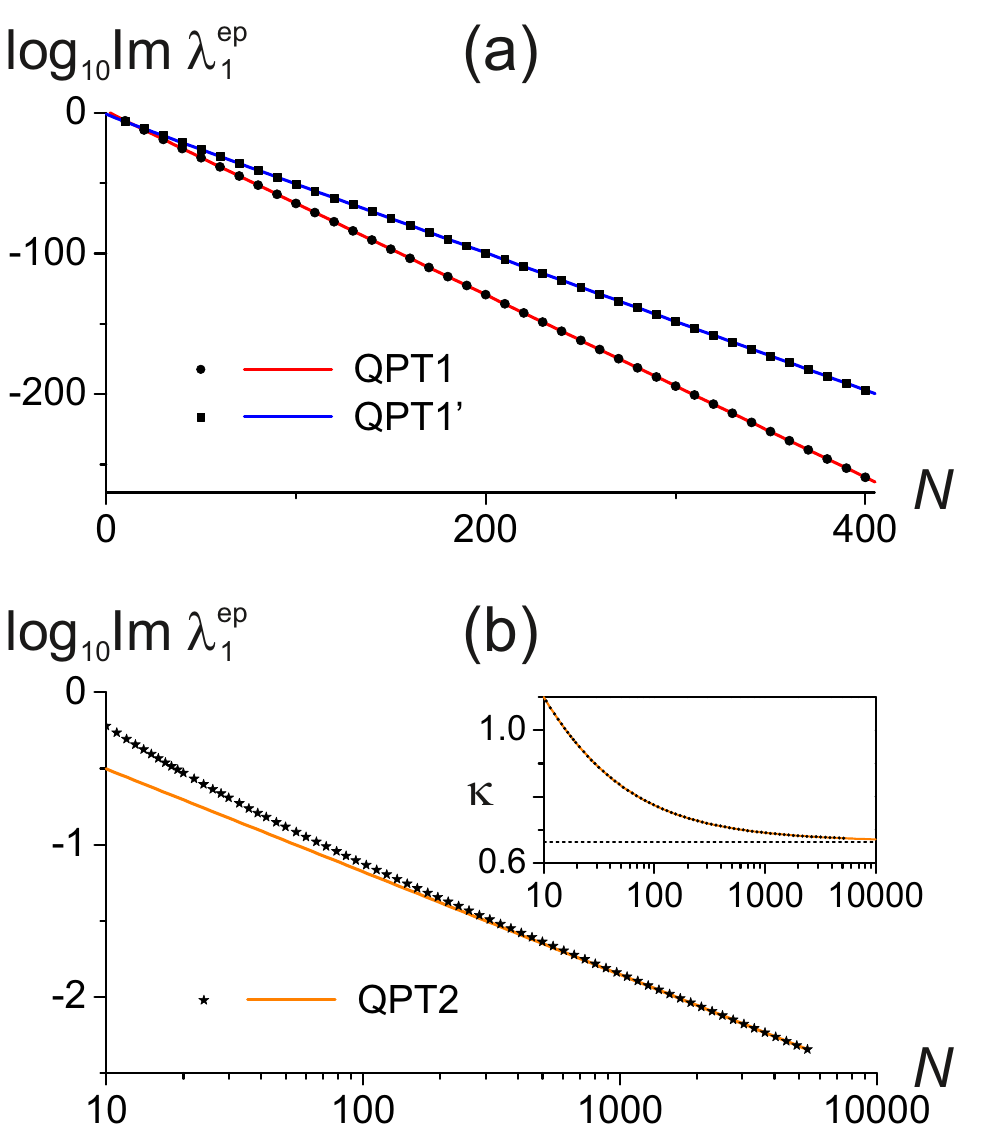}
\caption{
Evolution of the single \EP\ located closest to the \QPT\ critical point with increasing size $N$.
Panel (a): the first-order \QPT\ Lipkin Hamiltonians \eqref{HL1} and \eqref{HL1'} showing an exponential decrease of ${\rm Im}\vecb{\lambda}^{\rm ep}_1$. 
Panel (b): the second-order \QPT\ Lipkin Hamiltonian \eqref{HL2} with an algebraic decrease  of ${\rm Im}\vecb{\lambda}^{\rm ep}_1$.
A linear fit of the lin-log dependences in panel (a) yields ${\rm Im}\lambda^{\rm ep}_1\!\propto\!N^{-\zeta}e^{-\eta N}$, where $(\zeta,\eta)\!\doteq\!(0.52,1.49)$ for $\hat{H}^{\rm qpt1}$ and $(0.56,1.12)$ for $\hat{H}^{\rm qpt1'}$. 
The log-log dependence in panel (b) is consistent with ${\rm Im}\lambda^{\rm ep}_1\!\propto\!N^{-\kappa(N)}$, the evolution of $\kappa(N)$ being shown in the inset (an estimated asymptotic value deduced from calculation up to $N\approx 5400$ is $\kappa\!\doteq\!0.666$; the tilted line in the main graph is a linear fit through last three points).
}
\label{Fscal}
\end{figure}

The \EP-based distinction between the first- and second-order \QPTs\ can be formulated in a quantitative way by tracing the convergence of the nearest \EP\ to the critical point on the real axis with increasing size of the system.
This is presented in Fig.\,\ref{Fscal} for the above-studied Lipkin Hamiltonians.
We show the logarithm of ${\rm Im}\vecb{\lambda}^{\rm ep}_1$ (where index 1 is assigned to the closest \EP) as a function of $N$.
The horizontal scale is linear for the two first-order \QPTs\ in panel (a) and logarithmic for the second-order \QPT\ in panel (b), implying an exponential and roughly algebraic convergence of the nearest \EP\ to the critical point for the first- and second-order \QPT, respectively.
This means:
\begin{equation}
{\rm Im}\vecb{\lambda}^{\rm ep}_1\propto
\left\{\begin{array}{ll}
\exp(-\eta N\-\zeta\ln N) & {\rm for\ \QPT\,1,}\\
N^{-\kappa} & {\rm for\ \QPT\,2,}\\
\end{array}\right.
\label{appro}
\end{equation}
where $\eta,\zeta,\kappa$ are some positive constants.
We note that the log-log dependence in Fig.\,\ref{Fscal}(b) indicates a relatively slow convergence to the algebraic formula in \Eqref{appro}.
The exponent $\kappa$ exhibits a secondary dependence on $N$, but this dependence seems to have an asymptotic value $\lim_{N\to\infty}\kappa\approx\frac{2}{3}$ (see the inset of the figure and the line fitting the highest-$N$ points).

As follows from \Eqref{Edif}, there is a direct relation between the distance ${\rm Im}\vecb{\lambda}^{\rm ep}_1$ of the first \EP\ from the real axis and a spacing $\Delta_{21}\=E_2\-E_1$ between the two lowest states at $\lambda\={\rm Re}\vecb{\lambda}^{\rm ep}_1$. 
Indeed, the formula \eqref{appro} is consistent with the scaling of the critical spectra at the first- and second-order \QPT\ described in Appendix~\ref{SeApB} and in \Eqsref{Ec1} and \eqref{Ec2} below with a substitution $d\!\sim\!N$.
We have checked that for the first-order \QPT\ the relation between the exponential dependences in \Eqsref{appro} and \eqref{Ec1} is quantitative, yielding the same constants in the exponential. 
This holds not only for the binary avoided crossing of the lowest levels, but also for higher ones.

In contrast, the algebraic dependences in \Eqsref{appro} and \eqref{Ec1},  associated with the second-order \QPT, are related only in a qualitative sense.
The exponent $\kappa\!\approx\!\frac{2}{3}$ characterizing large-$N$ scaling of ${\rm Im}\vecb{\lambda}^{\rm ep}_1$ differs from the value $\frac{1}{3}$ that describes the scaling of the energy spacing $\Delta_{21}$ between the lowest levels.
This discrepancy can be attributed to the proximity of several \EPs\ involving the lowest energy levels to the second-order \QPT\ critical point, see  \Figref{FL2}(b).
The function ${\cal F}_{nn'}(\lambda)$ in \Eqref{Edif}, which hides the influence of the neighboring \EPs, cannot be assumed to vary slowly on the real axis, and the minimal spacing $\Delta_{21}$ is located at a certain $\lambda_0$ shifted away from $\lambda\={\rm Re}\vecb{\lambda}^{\rm ep}_1$.
Rewriting \Eqref{Edif} as $\Delta_{21}(\lambda_0)\=2{\cal F}_{21}(\lambda_0)[\left(\lambda_0\-{\rm Re}\vecb{\lambda}^{\rm ep}_1\right)^2\+{\rm Im}^2\vecb{\lambda}^{\rm ep}_1]^{1/2}$, we see that the large-$N$ scalings of $\Delta_{21}(\lambda_0)$ and ${\rm Im}\vecb{\lambda}^{\rm ep}_1$ need not be the same.

\section{Exceptional points for randomly perturbed critical Hamiltonians}
\label{SeRa}

In this Section, the linear Hamiltonian form \eqref{Hlin} is studied from a different perspective. 
The free term $\hat{H}(0)$ is associated with the {\em critical-point Hamiltonian\/} $\hat{H}^{\rm c1}$ or $\hat{H}^{\rm c2}$ of a first- or second-order \QPT\ taken from the model of Sec.\,\ref{SeLi}, while the interaction term $\hat{V}$ is considered as a {\em random matrix}.
We want to study to what extent the criticality of $\hat{H}^{{\rm c}k}$ represents a property independent of a particular model-specific Hamiltonian trajectory.
Will the critical properties of $\hat{H}^{{\rm c}k}$ be preserved even in this setup?
Does an arbitrary perturbation of a critical Hamiltonian show some universal features in the distribution of \EPs?
Note that analyses of linear Hamiltonians with a random interaction term were presented in Refs.\,\cite{Zir83,Sha17}, but only with a noncritical Hamiltonian $\hat{H}(0)$. 
Here we extend these studies by considering various forms of the free Hamiltonian and also different classes of random perturbations.

\subsection{Hamiltonian forms}

The full Hamiltonian ${\hat H}(\lambda)$ is expressed in the unperturbed eigenbasis, so that the free Hamiltonian is represented by a diagonal matrix
\begin{equation} 
{\hat H}(0)={\rm diag}\,\{E_1(0),E_2(0),\dots,E_d(0)\}
=\left\{\begin{array}{l}
\hat{H}^{\rm c1}\,,\\
\hat{H}^{\rm c2}\,,\\
\hat{H}^{\rm ho}\,,
\end{array}\right.
\label{Hfree}
\end{equation}
where $d$ is the dimension.
For ${\hat H}(0)\=\hat{H}^{\rm c1}$, that is for the critical Hamiltonian of the first-order \QPT, the energies $E_n(0)\=E_n^{\rm c1}$ are those of a parity-symmetric degenerate double-well Hamiltonian in one degree of freedom.
We employ a numerical spectrum of the Lipkin Hamiltonian \eqref{HL1} with $\lambda\=\lambda^{\rm c1}$.
The spectrum inside the wells consists of parity doublets, the separation of levels inside the doublet quickly decreasing with increasing $\hbar^{-1}\!\propto\!d$.
As shown in Appendix~\ref{SeApB}, the spacings between neighboring levels for $d\!\gg\!1$ and $n\!\ll\!d$ can be semiclassically approximated as:
\begin{equation}
E_{n+1}^{\rm c1}\-E_n^{\rm c1}\approx
\left\{\begin{array}{ll}
2\omega & {\rm for\ }n{\rm\ even,}\\
A_n\exp(-B_nd\-C_n\ln d) & {\rm for\ }n{\rm\ odd,}
\end{array}\right.
\label{Ec1}
\end{equation}
where $\omega$ is an average spacing, while $A_n,B_n,C_n$ are some positive constants.

For ${\hat H}(0)\=\hat{H}^{\rm c2}$, that is for a second-order \QPT, which is for one degree of freedom associated with the pure quartic oscillator, we use a numerical spectrum of the Lipkin Hamiltonian \eqref{HL2} at $\lambda\=\lambda^{\rm c2}$.
This spectrum for $d\!\gg$1 can be approximated by an explicit formula $E_n(0)\=E_n^{\rm c2}\approx\omega\,n^{4/3}d^{-1/3}$ (see Appendix~\ref{SeApB}), so:
\begin{equation}
E_{n+1}^{\rm c2}\-E_n^{\rm c2}\approx\frac{4\omega}{3}\left(\frac{n}{d}\right)^{\frac{1}{3}}
\label{Ec2}\,.
\end{equation}

Finally, to provide a comparison of the above critical cases with a noncritical one and to keep a link to the results of Refs.\,\cite{Zir83,Sha17}, we consider also the third choice of the free Hamiltonian, ${\hat H}(0)\=\hat{H}^{\rm ho}\!\propto\!\hat{J}_3$, which has an equidistant spectrum {\em {\`a} la} harmonic oscillator, hence $E_n(0)\=E_n^{\rm ho}\=\omega n$ and:
\begin{equation}
E_{n+1}^{\rm c2}\-E_n^{\rm c2}=\omega
\label{Eho}\,.
\end{equation}

The random interaction term $\hat{V}$ will be associated with three different classes of random matrix ensembles:
\begin{equation} 
{\hat V}=\left\{\begin{array}{l}
\hat{V}^{\rm diag}\,,\\
\hat{V}^{\rm full}\,,\\
\hat{V}^{\rm offd}\,.
\end{array}\right.
\label{Vrand}
\end{equation}
The first choice, $\hat{V}^{\rm diag}$, represents purely {\em diagonal matrices\/} with elements $V_{nn'}^{\rm diag}\=0$ for $n\!\neq\!n'$ and $V_{nn}^{\rm diag}$ being independent random variables with zero expectation value and variance $\sigma^2$.
We consider either the normal distribution ${\rm N}(0,\sigma^2)$ with $V_{nn}^{\rm diag}\in(-\infty,+\infty)$, or the rectangular distribution ${\rm R}(0,\sigma^2)$ on the interval $V_{nn}^{\rm diag}\in[-\sqrt{3}\sigma,+\sqrt{3}\sigma]$.
That is:
\begin{equation}
V_{nn'}^{\rm diag}:=\left\{\begin{array}{ll}\sim\Bigl\{\begin{array}{cc}{\rm R}(0,\sigma^2)\\{\rm N}(0,\sigma^2)\end{array}\Bigr\}&{\rm for\ }n\=n'\,,\\0&{\rm for\ }n\!\neq\!n'\,,\end{array}
\right.
\label{Vdiag}
\end{equation}
where $\sim$ means \uvo{taken from}.

The second choice of $\hat{V}$ corresponds to the classical {\em Gaussian orthogonal ensemble\/} (\GOE) \cite{Meh04}.
As the whole interaction matrix is completely filled, we call this case $\hat{V}^{\rm full}$.
The matrix elements are normally distributed independent random variables generated via the following prescription:
\begin{equation}
V_{nn'}^{\rm full}:=\left\{\begin{array}{ll}\sim{\rm N}(0,2\sigma^2)&{\rm for\ }n\=n'\,,\\\sim{\rm N}(0,\sigma^2)&{\rm for\ }n\!\neq\!n'\,.\end{array}
\right.
\label{Vfull}
\end{equation}

The third choice, named $\hat{V}^{\rm offd}$, is similar to the previous one except that the diagonal matrix elements of the \GOE\ interaction are fully erased.
So we have a strictly {\em offdiagonal matrices\/} generated as:
\begin{equation}
V_{nn'}^{\rm offd}:=\left\{\begin{array}{ll}0&{\rm for\ }n\=n'\,,\\\sim{\rm N}(0,\sigma^2)&{\rm for\ }n\!\neq\!n'\,.\end{array}
\right.
\label{Voffd}
\end{equation}
Note that this ensemble of interaction matrices is expected to yield results partly similar as the matrices taken from the Gaussian unitary ensemble (\GUE); see Ref.\,\cite{Sha17} where the analysis is done for $\hat{H}(0)\=\hat{H}^{\rm ho}$.
This is due to the fact that the absolute size of diagonal matrix elements in a complex-valued \GUE\ matrix is suppressed relative to the offdiagonal ones, in analogy to the extreme offdiagonal case studied here.

The above three classes of random ensemble can be seen as some representative scenarios of perturbing the free Hamiltonian and breaking its symmetries.
The diagonal ensemble \eqref{Vdiag} corresponds to perturbations preserving all the symmetries of the original Hamiltonian.
As we assume a nondegenerate spectrum, the interaction term must be diagonalized in the same basis as $\hat{H}(0)$.
For the full-matrix ensemble \eqref{Vfull}, the eigenbasis of the interaction Hamiltonian is identified with a random rotation of the unperturbed basis.
Indeed, the \GOE\ is built in such a way that any eigenbasis rotation has an equal probability, so the information on initial symmetries is completely lost.
Finally, the offdiagonal ensemble \eqref{Voffd} captures the situations in which the initial symmetries are violated in a  maximal way so that the probability of conserving the unperturbed basis is zero.
Null diagonal matrix elements of the interaction indicate that the result of $\hat{V}$ acting on any unperturbed eigenvector is perpendicular to this eigenvector; imagine as an example $\hat{H}(0)\propto\hat{J_3}$ (an initial magnetic field in the $z$-direction) and $\hat{V}\propto(\hat{J}_+\+\hat{J}_-)\propto\hat{J}_1$ (a perturbing magnetic field in the $x$-direction).

Each of the matrix ensembles $\hat{V}^{\rm diag}$, $\hat{V}^{\rm full}$ and $\hat{V}^{\rm offd}$ has a free parameter---the variance $\sigma^2$ in \Eqsref{Vdiag}, \eqref{Vfull} and  \eqref{Voffd}.
This parameter determines the dispersion of diagonal and/or offdiagonal matrix elements and also a overall \uvo{size} of the interaction term averaged over the ensemble.
Therefore, it competes with the outer control parameter $\lambda$ of the whole Hamiltonian \eqref{Hlin}.
To avoid this ambiguity, we normalize $\sigma^2$ to make the average size of $\hat{V}$ equal to the fixed size of $\hat{H}(0)$.
We use a {\em quadratic spread\/} $D_E$ of the spectrum $\{E_n\}_{n=1}^{d}$, here for the sake of generality taken complex:
\begin{equation}
D_E=\frac{1}{d\-1}\sum_{n=1}^{d}\left|E_n\-M_E\right|^2=\frac{{\rm Tr}\hat{H}\hat{H}^{\dag}}{d\-1}-\frac{{\rm Tr}\hat{H}{\rm Tr}\hat{H}^{\dag}}{d(d\-1)}
\,.\label{Edis}
\end{equation}
Operator $\hat{H}$, not necessarily Hermitian, represents the spectrum generating Hamiltonian and 
\begin{equation}
M_E=\frac{1}{d}\sum_{n=1}^{d}E_n=\frac{{\rm Tr}\hat{H}}{d}
\,,\label{Emea}
\end{equation}
is the mean value, a \uvo{center of mass} of the spectrum.
Note that $\sqrt{D_E}$ quantifies the size (an average diameter) of the \uvo{cloud} of complex eigenvalues $\{E_n\}_{n=1}^d$ and plays a similar role as an operator norm of $\hat{H}\-M_E$.
For instance, $\hat{H}\=\hat{H}^{\rm osc}$ yields $\sqrt{D_E}\approx\omega d/\sqrt{12}$ for $d\gg 1$, while a pure quartic oscillator $\hat{H}\=\hat{H}^{\rm c2}$ has $\sqrt{D_E}\approx\omega d/\sqrt{11.23}$.

The quadratic spread \eqref{Edis} can be evaluated for the spectra of both the free and interaction terms of the Hamiltonian.
The adjustment of parameter $\sigma$ is therefore performed so that an expectation value $\ave{D_V}$ of the quadratic spread 
\begin{equation}
D_V=\frac{{\rm Tr}\hat{V}^2}{d\-1}-\frac{{\rm Tr}^2\hat{V}}{d(d\-1)}
\label{Vdis}
\end{equation}
of the spectrum of the random perturbation $\hat{V}$ is set equal to the quadratic spread $D_E(0)$ of the spectrum of the free Hamiltonian $\hat{H}(0)$.
For the above classes of perturbation ensembles this means:
\begin{equation}
\sigma^2=\left\{\begin{array}{ll}
D_E(0)&{\rm for\ }\hat{V}^{\rm diag}\,,\\
D_E(0)/(d\+2)&{\rm for\ }\hat{V}^{\rm full}\,,\\
D_E(0)/d&{\rm for\ }\hat{V}^{\rm offd}\,.
\end{array}\right.
\label{sig}
\end{equation}
Note that $\sigma^2$ in the full and offdiagonal cases is reduced by a factor $\sim\!1/d$ with respect to the diagonal case; this is caused by widening of the spectrum of a nondiagonal matrix due to level repulsion.   
The normalization \eqref{sig} implies that the strongest competition between the free and interaction terms of Hamiltonian \eqref{Hlin} is expected in a vicinity of $\lambda\=1$.

This overall expectation is supported by an analysis  of the global spectral measures \eqref{Edis} and \eqref{Emea} for Hamiltonians with running parameter $\lambda$.
Their evaluation is performed in Appendix~\ref{SeApC}.
It turns out that a perturbation of an arbitrary free Hamiltonian by a single random matrix from either of the above ensembles induces immediate spectral redistributions within an interval, which is placed nearly symmetrically around $\lambda\=0$ and whose width is of the order of unity.  
Most of the avoided level crossings should take place within this interval of $\lambda$ and the associated complex \EPs\ should be located around.
This bulk expectation was for the $\hat{V}^{\rm full}$ ensemble confirmed in Ref.\,\cite{Zir83}, where the \EP\ distribution of a \GOE-perturbed regular Hamiltonian was first studied. 
In the following, we analyze the actual \EP\ distributions in the complex $\vecb{\lambda}$ plane for various choices of $\hat{V}$ and $\hat{H}(0)$.

\subsection{Distributions of exceptional points}

Let us study the distributions of \EPs\ associated with the three types of free Hamiltonian \eqref{Hfree} and the three classes of random interaction \eqref{Vrand}.
It is clear that each sample matrix $\hat{V}$ taken from any ensemble gives a particular arrangement of discrete \EPs\ in the plane $\vecb{\lambda}\in\mathbb{C}$.
We are however interested in {\em smoothed distributions\/} of \EPs, which are obtained by averaging over the whole ensemble of interaction terms of the given class (or, if performed numerically, over a sufficiently large number of samples).

We will see that the three random interaction ensembles $\hat{V}^{\rm diag}$, $\hat{V}^{\rm full}$ and $\hat{V}^{\rm offd}$ exhibit crucially different average distributions of \EPs. 
For the diagonal ensemble $\hat{V}^{\rm diag}$, all degeneracies must be trivially located along the line $\vecb{\lambda}=\lambda+i\,0$.
They represent unavoided level crossings, ordinary diabolic points, that arise from a fusion of complex conjugate pairs $\vecb{\lambda}^{\rm ep}_i$ and ${\vecb{\lambda}^{\rm ep}_i}^{*}$ of \EPs\ at a point $\lambda^{\rm dp}_i$ on the real axis (fusion of a pair of \EPs\ can in general produce either a \DP\ or a higher-order type of singularity).
For the full-matrix ensemble $\hat{V}^{\rm full}$, the \EPs\ are scattered in the whole complex plane.
It turns out that the ensemble-averaged \EP\ distribution for the \GOE\ perturbation is rotationally symmetric---depending just on $\abs{\vecb{\lambda}}$ after the full averaging \cite{Sha17}.
Finally, for the offdiagonal ensemble $\hat{V}^{\rm offd}$ the distribution of \EPs\ is located in regions closer to the imaginary axis.
So the succession $\hat{V}^{\rm diag}\tto\hat{V}^{\rm full}\tto\hat{V}^{\rm offd}$ captures a sampled view of a gradual move of \EPs\ in the complex $\vecb{\lambda}$ plane from the real axis towards the imaginary axis.

\begin{figure}[!t]
\includegraphics[width=\linewidth]{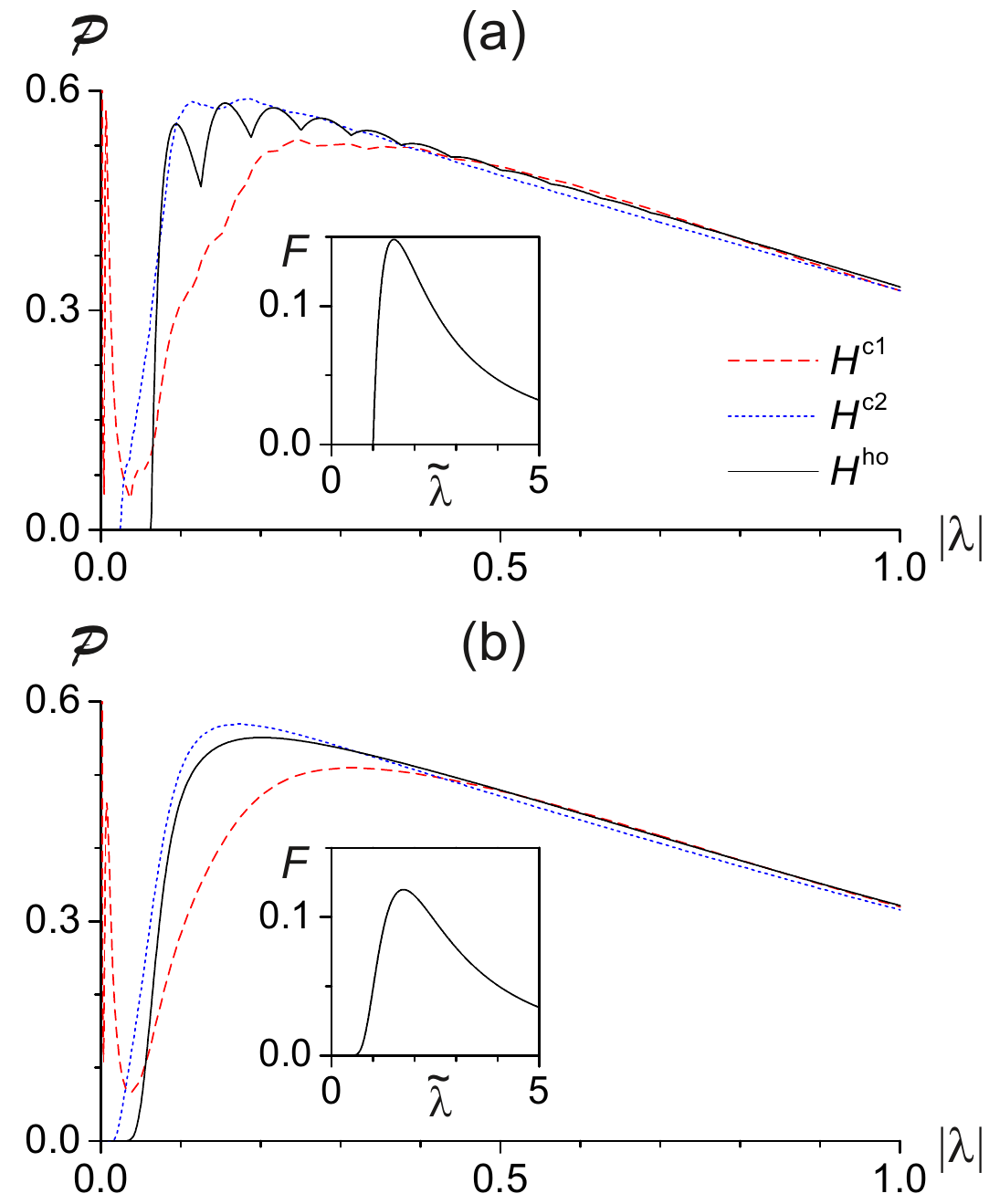}
\caption{
Ensemble-averaged distributions of \DPs\ (real crossings) along $|\lambda|\!=\!|{\rm Re}\vecb{\lambda}|$ for Hamiltonian \eqref{Hlin} with the diagonal random interaction $\hat{V}\!=\!\hat{V}^{\rm diag}$ from \Eqref{Vdiag} for $d\!=\!16$ ($N\!=\!15$).
The calculation was done via formula \eqref{Decro}. 
Panel (a) corresponds to the rectangular and panel (b) to the normal distribution of diagonal matrix elements; the functions $F$ from \Eqref{Fgr} are shown in the respective insets.
Individual curves show results for three unperturbed Hamiltonians: $\hat{H}(0)\!=\!\hat{H}^{\rm c1}$,  $\hat{H}^{\rm c2}$ and  $\hat{H}^{\rm ho}$.
}
\label{Fdiag}
\end{figure}

We start with the simplest diagonal case \eqref{Vdiag}.
It can be shown (see Appendix~\ref{SeApD}) that the distribution of crossings $\lambda^{\rm dp}_i$ along the real axis $\vecb{\lambda}\=\lambda\>0$ is given by a formula:
\begin{equation}
{\cal P}(\lambda)=\frac{2}{{\cal I}}\sum_{n=1}^{d}\sum_{n'=n+1}^{d}\frac{2V_0}{\Delta_{nn'}(0)}\,F\!\left(\lambda\,\frac{2V_0}{\Delta_{nn'}(0)}\right)
\label{Decro}\,.
\end{equation}
Here, $\Delta_{nn'}(0)=E_{n'}(0)\-E_n(0)$  are differences of unperturbed energies of the Hamiltonian $\hat{H}(0)$, and $F(\tilde{\lambda})$ is a certain function derived from the distribution $p(v_{n})$ of the diagonal matrix elements $v_n\=V^{\rm diag}_{nn}/V_0$.
The latter distribution is expressed with respect to an arbitrary interaction energy scale $V_0$. 
Information on a particular level pair $n,n'$ in each term of \Eqref{Decro} is then reduced just to a dimensionless \uvo{form factor} $\alpha_{nn'}\=2V_0/\Delta_{nn'}(0)$.
We choose a value 
\begin{equation}
V_0\=\sqrt{3}\,\sigma\=\sqrt{3D_E(0)}
\,,
\end{equation} 
which e.g. for a harmonic oscillator yields $2V_0\=\omega d$.
This setting guarantees that the interval $V_{nn}\in[-V_0,+V_0]$ covers 100\,\% of the available values for the rectangular distribution ${\rm R}(0,\sigma^2)$ and approximately 92\,\% of all values for the normal distribution ${\rm N}(0,\sigma^2)$.
For both these distributions the function $F(\tilde{\lambda})$ can be written explicitly:
\begin{equation}
F(\tilde{\lambda})\=\left\{\begin{array}{ll}
\Theta(\tilde{\lambda}\-1)\,\tilde{\lambda}^{-3}(\tilde{\lambda}\-1) & {\rm for\ }{\rm R}\!\left(0,\sigma^2\right)\,,\\
(3/\pi)^{\frac{1}{2}}\tilde{\lambda}^{-2}\exp(-3\tilde{\lambda}^{-2}) & {\rm for\ }{\rm N}\!\left(0,\sigma^2\right)\,,
\end{array}\right.
\label{Fgr}
\end{equation}
where $\Theta(x)$ stands for a step function ($\Theta\=0$ for $x\<1$ and $\Theta\=1$ for $x\!\geq\!1$).
These dependences are displayed in the insets of Fig.\,\ref{Fdiag}.
The form of $F(\tilde{\lambda})$ for a general distribution of diagonal matrix elements and the derivation of the above formulas is presented in Appendix~\ref{SeApD}.

The formula \eqref{Decro} is normalized to yield a unit integral over the range $\lambda\in[0,+\infty)$, as can be checked for both specific functions in \Eqref{Fgr}.
The crossings are distributed symmetrically with respect to $\lambda\=0$, so we can replace ${\cal P}(\lambda)$ by ${\cal P}(\abs{\lambda})$.
As the whole range $\lambda\in(-\infty,+\infty)$ contains a total number of ${\cal I}\=d(d\-1)/2$ crossings, the dimension-dependent density of crossings ${\cal D}(\abs{\lambda})$ is given by \Eqref{Decro} without the prefactor.

The ensemble-averaged distributions of crossings obtained from the formula \eqref{Decro} for $\hat{H}(0)\=\hat{H}^{\rm c1}$, $\hat{H}^{\rm c2}$ and $\hat{H}^{\rm ho}$ are displayed in the main panels of Fig.\,\ref{Fdiag} for a moderate dimension $d\=16$.
Panel (a) corresponds to the rectangular distribution of diagonal matrix elements, panel (b) to the normal distribution.
Although the rectangular distribution yields a sharper form of the function $F$ than the normal distribution (see the insets), both cases result in similar overall dependences ${\cal P}(\abs{\lambda})$.
We observe that if the free Hamiltonian is taken at the first-order \QPT, $\hat{H}(0)\=\hat{H}^{\rm c1}$, the distribution has a sharp peak at very small values of $|\lambda|$.
This is a direct consequence of the nearly degenerate parity doublets associated with the reflection-symmetric critical Hamiltonian resulting from \Eqref{HL1}.
As the spacings $\Delta_{nn'}(0)$ between the doublet states decreases exponentially with dimension $d$, see \Eqref{Ec1}, the peak quickly converges to $\abs{\lambda}\=0$ with $d\tto\infty$.
In this limit, the width of the peak vanishes and its height diverges.
Such an effect is not present if $\hat{H}(0)$ is associated with the critical Hamiltonian $\hat{H}^{\rm c2}$, or with a harmonic oscillator $\hat{H}^{\rm ho}$.
Nevertheless, the second-order \QPT\ Hamiltonian $\hat{H}^{\rm c2}$ still shows a clearly distinguished shift of ${\cal P}(\abs{\lambda})$ towards smaller values of $\abs{\lambda}$ in comparison with $\hat{H}^{\rm ho}$.
This is obviously a consequence of the cumulation of levels in the pure quartic oscillator near the ground state, see \Eqref{Ec2}.

\begin{figure}[!t]
\includegraphics[width=\linewidth]{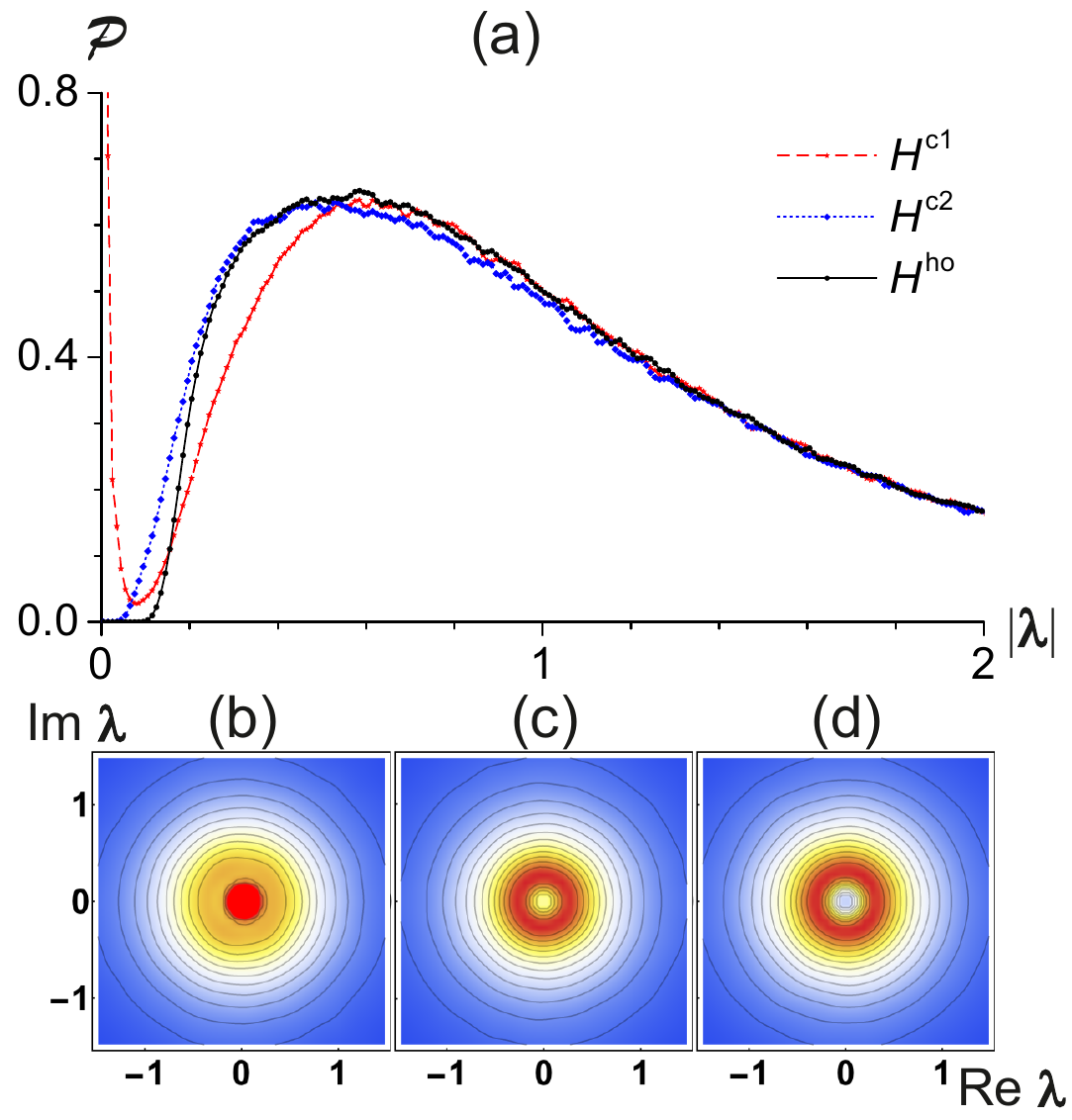}
\caption{
Ensemble-averaged distributions of \EPs\ for Hamiltonian \eqref{Hlin} with the \GOE\ random interaction $\hat{V}\!=\!\hat{V}^{\rm full}$ from \Eqref{Vfull} for $d\!=\!16$ ($N\!=\!15$).
Panel (a): \EP\ distribution as a function of the absolute value $|\vecb{\lambda}|$; the three curves correspond to the three choices of the free Hamiltonians \eqref{Hfree}.
Lower panels: \EP\ distributions in the whole complex plane for $\hat{H}(0)\!=\!\hat{H}^{\rm c1}$ (b), $\hat{H}^{\rm c2}$ (c) and $\hat{H}^{\rm ho}$ (d).
The distribution was calculated from $\sim$8400 random-matrix realizations ($\sim$$10^6$ complex-conjugate pairs of \EPs).
 }
\label{Ffull}
\end{figure}

\begin{figure}[!t]
\includegraphics[width=\linewidth]{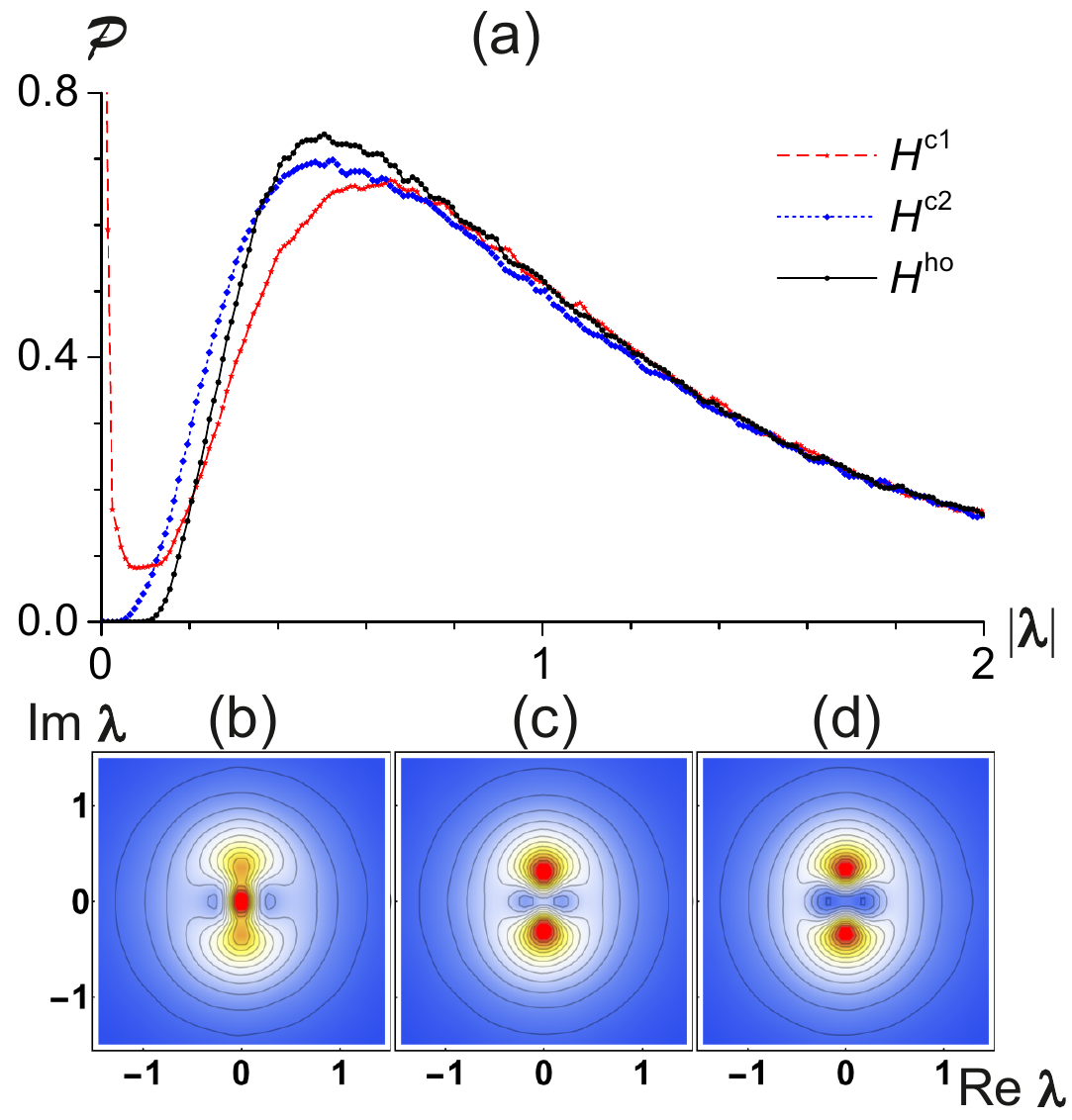}
\caption{
The same as in Fig.\,\ref{Ffull}, but for the offdiagonal random interaction ensemble $\hat{V}\!=\!\hat{V}^{\rm offd}$ from \Eqref{Voffd}.
}
\label{Foffd}
\end{figure}

Let us proceed to the analysis of full and offdiagonal interaction matrices \eqref{Vfull} and \eqref{Voffd}.
The ensemble-averaged distributions of $\vecb{\lambda}^{\rm ep}_i$  in the complex plane for these Hamiltonians are shown in \Figsref{Ffull} and \ref{Foffd}, respectively.
The lower panels in both figures show the distribution of \EPs\ in the whole complex $\vecb{\lambda}$ plane for $\hat{H}(0)$ associated with the first-order \QPT\ Hamiltonian $\hat{H}^{\rm c1}$ (panel~b), the second-order \QPT\ Hamiltonian $\hat{H}^{\rm c2}$ (panel~c), and for the harmonic oscillator $\hat{H}^{\rm ho}$ (panel~d).
Panel (a) in both figures depicts the distributions ${\cal P}(\abs{\vecb{\lambda}})$ of the absolute values $|\vecb{\lambda}^{\rm ep}_i|$ connected with the complex $\vecb{\lambda}^{\rm ep}_i$ distributions in the lower panels.
The distributions in the upper panels are normalized in the same way as those in \Figref{Fdiag}, i.e., to a unit integral over the whole range $|\vecb{\lambda}|\in[0,\infty)$.

As seen in the lower panels of \Figref{Ffull}, the ensemble-averaged distributions of \EPs\ for the full \GOE\ interaction matrix $\hat{V}^{\rm full}$ show a perfect rotational symmetry around the origin of the $\vecb{\lambda}$ plane for any choice of $\hat{H}(0)$.
This feature, which is violated for any departure from the \GOE\ class of perturbation, was recently discussed in Ref.\,\cite{Sha17}, noting that no obvious source of the symmetry has been identified so far.
In contrast, all \EP\ distributions for the offdiagonal interaction ensemble $\hat{V}^{\rm offd}$ in the lower panels of \Figref{Foffd} show a strong redistribution of \EPs\ towards the imaginary axis in the $\vecb{\lambda}$ plane.
A similar but less pronounced feature was observed for complex \GUE\ interaction matrices \cite{Sha17}; see the discussion below \Eqref{Voffd}.

Despite the significant differences between the entire $\vecb{\lambda}^{\rm ep}_i$ distributions for $\hat{V}\=\hat{V}^{\rm diag}$, $\hat{V}^{\rm full}$ and $\hat{V}^{\rm offd}$, the corresponding distributions ${\cal P}(|\vecb{\lambda}|)$ of absolute values $|\vecb{\lambda}^{\rm ep}_i|$ for a fixed $\hat{H}(0)$ do not differ too much, see the upper panels in \Figsref{Ffull} and \ref{Foffd}, and both panels in \Figref{Fdiag}.
One may notice that the nondiagonal ensembles in \Figsref{Ffull}(a) and \ref{Foffd}(a) in comparison with the diagonal ensembles in \Figref{Fdiag} yield the peak area of ${\cal P}(|\vecb{\lambda}|)$ slightly shifted to larger values of $|\vecb{\lambda}|$ and simultaneously suppress the long-range tail of ${\cal P}(|\vecb{\lambda}|)$.
This is a consequence of correlations caused by nondiagonal matrix elements in both nondiagonal ensembles.
The uncorrelated diagonal elements of $\hat{V}^{\rm diag}$ show no repulsion and therefore lead to undelayed crossings of fast-converging levels as well as to very late crossings of levels with similar slopes.
In contrast, the nondiagonal ensembles $\hat{V}^{\rm full}$ and $\hat{V}^{\rm offd}$ suppress crossings with both small and large values of $|\vecb{\lambda}|$.
Except these differences, the ${\cal P}(|\vecb{\lambda}|)$ distributions for various interaction classes look qualitatively similar.

On the other hand, the ${\cal P}(|\vecb{\lambda}|)$ distributions differ considerably for various choices of the free Hamiltonian $\hat{H}(0)$. 
Taking the harmonic-oscillator case $\hat{H}(0)\=\hat{H}^{\rm ho}$ as a reference, we see that both \QPT\ critical Hamiltonians $\hat{H}^{{\rm c}1}$ and $\hat{H}^{{\rm c}2}$ shift the distributions towards lower values of $\abs{\vecb{\lambda}}$.
While the second-order critical Hamiltonian $\hat{H}^{\rm c2}$ leads only to a small but noticeable shift, the first-order critical Hamiltonian $\hat{H}^{\rm c1}$ creates a sharp peak of ${\cal P}(\abs{\vecb{\lambda}})$ at nearly zero values of $\abs{\vecb{\lambda}}$.
These conclusions hold for all interaction ensembles.

\begin{figure}[!t]
\includegraphics[width=\linewidth]{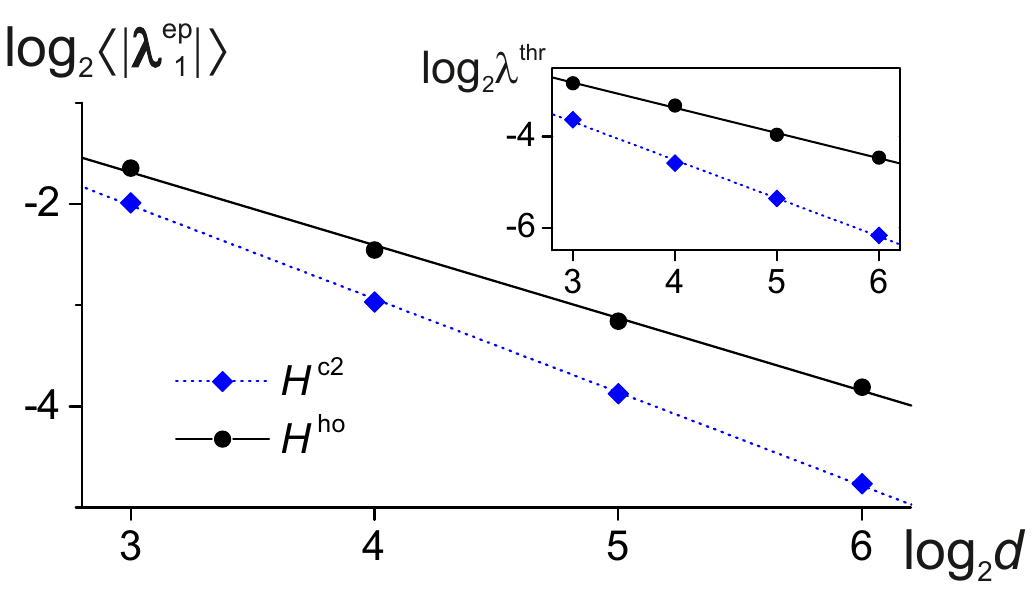}
\caption{
Behavior of \EPs\ near $\vecb{\lambda}\!=\!0$ for $\hat{H}(0)\!=\!\hat{H}^{\rm c2},\hat{H}^{\rm ho}$ and $\hat{V}\!=\!\hat{V}^{\rm full}$ with increasing dimension  ($d$=8,16,32,64).
The main panel shows the ensemble average $\ave{|\vecb{\lambda}_1^{\rm ep}|}$ of the deviation of the closest \EP\ from the origin.
The inset shows the value $\lambda^{\rm thr}$ determined as the lowest value $|\vecb{\lambda}_i^{\rm ep}|$ in a sample of $\approx$\,$10^6$ generated \EPs.
Linear fits (lines) indicate a faster convergence of both $\ave{|\vecb{\lambda}_1^{\rm ep}|}$ and $\lambda^{\rm thr}$ to zero for $\hat{H}^{\rm c2}$ than for $\hat{H}^{\rm ho}$.
 }
\label{Fthre}
\end{figure}

The explanation of this phenomenon is the same as for the diagonal ensemble:
The critical Hamiltonians $\hat{H}^{{\rm c}1}$ and $\hat{H}^{{\rm c}2}$ contain pairs or clusters of mutually close energy levels, therefore some of their \EPs\ (or \DPs) are located close to the origin $\vecb{\lambda}\=0$.
For the first-order \QPT\ this results in the peak exponentially approaching to the origin with increasing $d$, for the second-order \QPT\ there is only a certain shift in comparison with noncritical free Hamiltonians. 

To illustrate the latter difference in a more qualitative way, we follow in \Figref{Fthre} the evolution of \EPs\ located nearest to the origin $\vecb{\lambda}\=0$ with dimension $d$ ranging from 8 to 64.
Only two free Hamiltonians are compared, $\hat{H}^{\rm c2}$ and $\hat{H}^{\rm ho}$, while the interaction is taken as $\hat{V}^{\rm full}$.
The main panel shows the quantity $\ave{|\vecb{\lambda}_1^{\rm ep}|}$, which is the absolute value of the closest-to-origin \EP\ at $\vecb{\lambda}_1^{\rm ep}$ averaged over the whole  interaction ensemble.
Clearly, the average distance of the closest \EP\ from $\vecb{\lambda}\=0$ decreases with $d$ faster for the second-order \QPT\ Hamiltonian than for the harmonic-oscillator Hamiltonian.
Linear fits of the log-log dependences result in the estimates $\ave{|\vecb{\lambda}_1^{\rm ep}|}\sim d^{-0.93}$ for $\hat{H}^{\rm c2}$  and $\sim d^{-0.72}$ for $\hat{H}^{\rm ho}$.
We note that the dispersions of the $|\vecb{\lambda}_1^{\rm ep}|$ distributions in the ensemble of random interactions are relatively large for low dimensions, but they quickly decrease with increasing $d$.

Even a stronger effect can be seen in the inset of \Figref{Fthre}, where we show a \uvo{threshold} value $\lambda^{\rm thr}$ obtained as the closest-to-origin \EP\ in the whole sample of all generated \EPs.
Though this quantity depends on the size of the sample, its scaling with $d$ captures the behavior of the low \uvo{edge} of the ${\cal P}(|\vecb{\lambda}|)$ distributions in \Figref{Ffull}.
The fits indicate that $\lambda^{\rm thr}\sim d^{-0.84}$ for $\hat{H}^{\rm c2}$  and $\sim d^{-0.55}$ for $\hat{H}^{\rm ho}$.

A similar quantitative treatment of the first-order \QPT\ Hamiltonian $\hat{H}^{\rm c1}$ is hindered by some numerical problems in manipulation with nearly degenerate energy doublets for large dimensions in the nondiagonal setting.
Nevertheless, the insight gained from the diagonal crossing formula \eqref{Decro} leads us to anticipate that for $\hat{V}\=\hat{V}^{\rm full}$ and $\hat{V}^{\rm offd}$, in analogy with $\hat{V}^{\rm diag}$, the low-$|\vecb{\lambda}|$ peak of ${\cal P}(|\vecb{\lambda}|)$ associated with $\hat{H}^{\rm c1}$ tends to form a $\delta$-function type of singularity at $\vecb{\lambda}\=0$ in the asymptotic regime $d\tto\infty$.

\section{Conclusions}
\label{SeCo}

We have studied distributions of exceptional points near quantum phase transitions of the first and second order.
Initially, we have focused on some examples of \QPTs\ of both types in the simple Lipkin model.
We have seen that as the size parameter of the model increases, some \EPs\ converge to the critical point on the real axis of the complex $\vecb{\lambda}$ plane.
The convergence is exponential and algebraic for the first- and second-order \QPT, respectively.
This reflects, on one side, an exponential decrease of the tunneling probability between two wells of the first-order critical Hamiltonian and, on the other, an algebraic accumulation of energy levels near the ground state for the second-order critical Hamiltonian.
The first-order \QPT\ is connected with a single pair of \EPs\ that gets much closer to the real axis than the others, so that for a finite size it shows up as a sharp avoided crossing of a single pair of levels.
In contrast, the second-order \QPT\ is a more \uvo{collective} phenomenon in the sense that the properties of the ground state are simultaneously affected by several \EPs\ located at comparable distances from the real axis.  

In the second part of the paper, we have extended our analysis beyond the Lipkin model, considering critical first- and second-order \QPT\ Hamiltonians perturbed by various classes of random interactions (interaction ensembles).
We have seen that after a convenient normalization, the interaction term of any kind causes immediate [taking place for $\lambda\lesssim{\cal O}(1)$, independently of dimension] dispersion of the spectrum regardless of the unperturbed Hamiltonian.
However, it turned out that an initial stage of the dispersion process, governed by the ensemble-averaged distribution of \EPs\ close to $\vecb{\lambda}\=0$, carries a decisive information on the \QPT\ type.
In particular, for the first- and second-order \QPT, respectively, some of the \EPs\ either exponentially accumulate at, or algebraically converge to the $\vecb{\lambda}\=0$ point associated with the unperturbed critical Hamiltonian.  
These findings make us conclude that the distribution of \EPs\ represents a strong signature of quantum criticality that enables an unambiguous discrimination between the first- and higher-order critical Hamiltonians independently from a particular model parametrization.

Based on the \EP-related studies presented in Refs.~\cite{Cej07,Sin17}, a similar analysis like here can be performed also for excited-state \QPTs, i.e., nonanalyticities affecting higher energy levels in the spectrum \cite{Cej06,Cap08}.
As the classification of those transitions is entirely different from the classification of the ground-state \QPTs\ \cite{Str16}, the present results cannot be directly extrapolated to them.

Properties of the \EP\ distributions near the ground- or excited-state \QPTs\ may have important consequences for the superradiance phenomenon in open quantum systems---a sudden separation of short- and long-living states with an increasing transition rate into a common decay channel \cite{Sok92,Jun99,Aue11}.
This phenomenon is intimately connected with the location of \EPs\ in the non-Hermitian extension of the Hamiltonian, hence shall be sensitive to the above-studied properties. 
These issues will be addressed in our forthcoming work.

\section*{Acknowledgments}
This work was supported by the Czech Science Foundation under project no. P203-13-07117S.

\appendix

\section{Non-Hermitian extension and exceptional points}
\label{SeApA}

Here we outline some elementary properties of the eigensolutions of Hamiltonian \eqref{Hlin} with parameter $\lambda$ extended to $\vecb{\lambda}\in\mathbb{C}$.
We assume $\hat{H}(0)$ and $\hat{V}$ being incompatible real symmetric matrices of dimension $d$.
For ${\rm Re}\vecb{\lambda}\neq 0$, the Hamiltonian $\hat{H}(\vecb{\lambda})$ is represented by a non-Hermitian complex symmetric matrix satisfying $[\hat{H}(\vecb{\lambda}),\hat{H}(\vecb{\lambda})^{\dag}]\neq 0$, which means that it is not unitarily diagonalizable \cite{Hor85}. 
There exist $d$ complex eigenvalues $\{E_n(\vecb{\lambda})\}_{n=1}^d$ found as roots of the characteristic polynomial (due to the above constraints symmetric under the complex conjugation of $\vecb{\lambda}$). 
If all eigenvalues are mutually different, the Hamiltonian is diagonalized with the aid of a biorthogonal system of left and right eigenvectors $\bra{\psi_n^{\rm L}(\vecb{\lambda})}$ and $\ket{\psi_n^{\rm R}(\vecb{\lambda})}$, which are related by matrix transposition (instead of full Hermitian conjugation).
If $\hat{S}^{\rm L}(\vecb{\lambda})$ is a matrix whose rows are the left eigenvectors, $\hat{S}^{\rm R}(\vecb{\lambda})$ a matrix with columns formed by the right eigenvectors, and $\hat{D}(\vecb{\lambda})\!\equiv\!{\rm diag}\{E_1(\vecb{\lambda}),...,E_d(\vecb{\lambda})\}$, the diagonalization can be expressed as:
\begin{equation}
\hat{S}^{\rm L}(\vecb{\lambda})\hat{H}(\vecb{\lambda})\hat{S}^{\rm R}(\vecb{\lambda})=\hat{S}^{\rm L}(\vecb{\lambda})\hat{S}^{\rm R}(\vecb{\lambda})\hat{D}(\vecb{\lambda})
\label{leri}\,.
\end{equation}
Since the biorthogonality $\scal{\psi_n^{\rm L}(\vecb{\lambda})}{\psi_{n'}^{\rm R}(\vecb{\lambda})}\=\delta_{nn'}$ implies that $\hat{S}^{\rm L}(\vecb{\lambda})\hat{S}^{\rm R}(\vecb{\lambda})\=\hat{I}$, with $\hat{I}$ denoting the identity, \Eqref{leri} represents an ordinary (though nonunitary) similarity transformation of $\hat{H}(\vecb{\lambda})$ to the diagonal form.

A more difficult situation is encountered if $m\geq 2$ of the eigenvalues $\{E_n(\vecb{\lambda})\}_{n=1}^d$ coincide.
Consider for the sake of simplicity a single $m\=2$ degeneracy $E_n(\vecb{\lambda})\=E_{n'}(\vecb{\lambda})$ at a particular value of $\vecb{\lambda}$ (the same degeneracy appears also at the complex-conjugate value).
The degeneracy may be a diabolic point, in which the complex dependences $E_n(\vecb{\lambda})$ and $E_{n'}(\vecb{\lambda})$ form a conical intersection just as in the Hermitian case with two real parameters \cite{Ber84}. 
This would leave the above-outlined diagonalization procedure intact, preserving two left-right pairs biorthogonal eigenvectors associated with both levels at the degeneracy point.
However, a more natural scenario is that the degeneracy represents a true branch point in the sense of complex analysis---that is an exceptional point $\vecb{\lambda}^{\rm ep}$ in the terminology initiated in Ref.\,\cite{Kat66}.
At this point, two Riemann sheets of a multivalued function $E(\vecb{\lambda})$ containing eigenvalues of $\hat{H}(\vecb{\lambda})$ are interconnected.
In that case, the diagonalization \eqref{leri} fails since at the \EP\ both levels have only a single pair of eigenvectors satisfying the selforthogonality condition $\scal{\psi_n^{\rm L}(\vecb{\lambda}^{\rm ep}_i)}{\psi_{n}^{\rm R}(\vecb{\lambda}^{\rm ep}_i)}\=0$.
A similarity transformation turns the Hamiltonian into the Jordan form with a nontrivial block
\begin{equation}
J_{nn'}(\vecb{\lambda}^{\rm ep}_i)=\left(\begin{smallmatrix}E_n(\vecb{\lambda}^{\rm ep}_i)&1\\0&E_n(\vecb{\lambda}^{\rm ep}_i)\end{smallmatrix}\right)
\end{equation}
on the diagonal \cite{Hor85}.

The behavior of complex energies $E_n(\vecb{\lambda})$ near an \EP\ is described by the so-called Puiseux expansion \cite{Kat66,Kno47}.
For an $m$-fold \EP, the expansion is written in terms of fractional powers $(\vecb{\lambda}\-\vecb{\lambda}^{\rm ep}_i)^{k/m}$, with $k\=1,2,...$ 
It holds for $\abs{\vecb{\delta}}\equiv\abs{\vecb{\lambda}\-\vecb{\lambda}^{\rm ep}_i}\<R$, where $R$ is the distance to the nearest \EP\ related to any of the $m$ levels involved in the \EP\ studied.
Starting at the Riemann sheet associated with an arbitrary level involved in the degeneracy and completing $m$ loops around $\vecb{\lambda}^{\rm ep}_i$, one returns to the original point after visiting Riemann sheets of all the other levels.
Therefore, an enumeration of levels in the complex spectrum is possible only locally.
We note that almost all non-Hermitian degeneracies of a generic Hamiltonian \eqref{Hlin} are of the $m\=2$ EP type.
While unlikeliness of the \DP\ degeneracies in the complex-parameter domain is connected with the necessity to delete all fractional-power terms in the Puiseux expansion, the suppression of $m\>2$ \EPs\ follows from a higher number of constraints needed for their occurrence.

Near an $m\=2$ \EP\ involving general levels $n$ and $n'$ the Puiseux expansion reads as:
\begin{equation}
\begin{array}{rl}
E_n(\vecb{\lambda})&=E_n(\vecb{\lambda}^{\rm ep}_i)\+\sum\limits_{k=1}^{\infty}a_k\left(\vecb{\lambda}\-\vecb{\lambda}^{\rm ep}_i\right)^{k/2},\\
E_{n'}(\vecb{\lambda})&=E_{n'}(\vecb{\lambda}^{\rm ep}_i)\+\sum\limits_{k=1}^{\infty}a_k(-)^k\left(\vecb{\lambda}\-\vecb{\lambda}^{\rm ep}_i\right)^{k/2},
\end{array}
\label{EPEn}
\end{equation}
where $E_n(\vecb{\lambda}^{\rm ep}_i)\=E_{n'}(\vecb{\lambda}^{\rm ep}_i)$ and $a_k\in\mathbb{C}$ stand for expansion coefficients.
Very close to the \EP, the lowest term dominates, yielding $(E_n\-E_{n'})\approx 2a_1\sqrt{\vecb{\delta}}$, which is not analytic.
However, one can introduce a function \cite{Hei91}
\begin{equation}
{\cal F}_{nn'}(\vecb{\lambda})\=\frac{E_{n}(\vecb{\lambda})\-E_{n'}(\vecb{\lambda})}{2\sqrt{(\vecb{\lambda}\-\vecb{\lambda}^{\rm ep}_i)(\vecb{\lambda}\-{\vecb{\lambda}^{\rm ep}_i}^{*})}}
\=\sum_{l=0}^{\infty}a_{2l+1}\frac{(\vecb{\lambda}\-\vecb{\lambda}^{\rm ep}_i)^l}{\sqrt{\vecb{\lambda}\-{\vecb{\lambda}^{\rm ep}_i}^{*}}}
\label{Ffunc},
\end{equation}
which is regular within the whole disc of radius $R$ around $\vecb{\lambda}^{\rm ep}_i$.
If real axis of $\vecb{\lambda}$ intersects this disc, the function \eqref{Ffunc} describes the real energy dependences $E_n(\lambda)$ and $E_{n'}(\lambda)$ on the corresponding interval.
In this way we derive the avoided-crossing formula \eqref{Edif}.

\section{Semiclassical approximations of critical spectra}
\label{SeApB}

We sketch the derivation of the approximate level spacing formulas \eqref{Ec1} and \eqref{Ec2} for critical Hamiltonians $\hat{H}^{\rm c1}$ and $\hat{H}^{\rm c2}$.
They are based on the semiclassical quantization condition:
\begin{equation}
S(E_n)\equiv\oint_{E=E_n}\!\!\!\! dx\,p=2\pi\hbar\,\left(n\-\tfrac{1}{2}\right)
\label{Bohr}\,,
\end{equation}
where $p$ is the momentum at coordinate $x$ for a given energy level $E_n$ enumerated by $n\=1,2,..$ \cite{Chi14}.
The second-order critical Hamiltonian following from \Eqref{HL2cl} is approximated for low energies by a pure quartic oscillator $H^{\rm c2}\approx Ap^2\+Bx^4$, where $A,B$ are constants. 
The integral in \Eqref{Bohr} then reads as $S(E_n)\=2IA^{-1/2}B^{-1/4}E_n^{3/4}$ with $I\=\int_{-1}^{+1}dx\sqrt{1\-x^4}$. 
Hence we get $E_n\!\approx\!Cn^{4/3}$, where the constant $C$ can be expressed through an average spacing $\omega\!\approx\!E_d/d$ as $C\=\omega d^{-1/3}$. 
This yields \Eqref{Ec2}.

The semiclassical spectrum for the first-order critical Hamiltonian is derived for a parity-symmetric double-well system, e.g. that from \Eqref{HL1cl}.
The formula \eqref{Bohr} is applied in both wells separately, yielding each level $E_n$ two-fold degenerate (for energies bellow the barrier).
The eigenstates $\psi_{n\pm}(x)$ with parity $\pm$ are obtained by imposing the conditions $\psi_{n+}(0)\=0$ and $\frac{d}{dx}\psi_{n-}(0)\=0$.
This leads to the following approximate expression for the corresponding energies $E_{n\pm}$: 
\begin{equation}
E_{n-}\!\-E_{n+}\approx\frac{8\hbar\exp\left(-\tfrac{T}{\hbar}\right)}{S'}\biggr|_{E=E_n}
\label{eveodd2},
\end{equation}
where $T\=\smallint dx\,|p|$ (integral taken across the barrier separating both wells) is related to the semiclassical tunneling probability $P\!\approx\exp(-2T/\hbar)$, while $S'\=\frac{\partial}{\partial E}S$ \cite{Chi14}.
Hence we obtain the second line of \Eqref{Ec1}.
The first line results from a harmonic approximation of states inside the wells and from the neglect of the parity-doublet spacings \eqref{eveodd2} relative to spacings of equal-parity states.

\section{Global properties of the $\vecb{\hat{H}(\lambda)}$ spectrum}
\label{SeApC}

We look at the mean value \eqref{Emea} and the quadratic spread \eqref{Edis} of the entire spectrum of $\hat{H}(\vecb{\lambda})$ with a general interaction $\hat{V}$, and particularly at the statistical features of these quantities if $\hat{V}$ is taken from the random ensembles \eqref{Vrand}.
Assuming an arbitrary Hamiltonian of the form \eqref{Hlin}, the spectral mean value is trivially given by:
\begin{equation}
M_E(\vecb{\lambda})=M_E(0)+\vecb{\lambda}M_V
\,,\label{Emeala}
\end{equation}
where $M_V\={\rm Tr}\hat{V}/d$.
This demonstrates a linear dependence of the \uvo{center of mass} of the spectrum on $\vecb{\lambda}\in{\mathbb C}$.
Similarly, for the quadratic spread we obtain:
\begin{eqnarray}
D_E(\vecb{\lambda})=&&D_E(0)+{\rm Re}\vecb{\lambda}\underbrace{\frac{2d}{d\-1}\left[M_{HV}(0)\-M_E(0)M_V\right]}_{K}
\nonumber\\[-8pt]
&&+|\vecb{\lambda}|^2 D_V
,\label{Edisla}
\end{eqnarray}
where $M_{HV}(0)\={\rm Tr}(\hat{H}(0)\hat{V})/d$ is the spectral mean value of a Hermitian operator $(\hat{H}(0)\hat{V}\+\hat{V}\hat{H}(0))/2$ and $D_V$ is defined in \Eqref{Vdis}. 
This shows a quadratic dependence of the quadratic spread of the spectrum on $\vecb{\lambda}$ \cite{Cej09}.
For ${\rm Im}\vecb{\lambda}\=0$, the formula \eqref{Edisla} defines a parabola with a minimum at $\lambda_0\=-K/2D_V$.
At this point, the real spectrum becomes maximally compressed, its quadratic spread being equal to $D_E(\lambda_0)\=D_E(0)\-K^2/4D_V$.
Main structural changes in the Hamiltonian eigenstates due to the competition between free and interaction terms take place in a $\Delta\lambda\approx\sqrt{D_E(\lambda_0)/D_V}$
vicinity of $\lambda_0$.
In contrast, for $|\lambda-\lambda_0|$ much larger, the interaction term becomes dominant, so the spectrum just linearly expands and the eigenvectors freeze up (for a fixed realization of $\hat{V}$).

For $\hat{V}$ associated with any of the ensembles \eqref{Vdiag}, \eqref{Vfull} or \eqref{Voffd}, the coefficients $M_V,D_V$ and $K$ in \Eqsref{Emeala} and \eqref{Edisla} are statistical variables.
Their expectation values are easy to calculate:
\begin{equation}
\ave{M_V}=0
\,,\quad
\ave{D_V}=D_E(0)
\,,\quad
\ave{K}=0
\,,\label{AvAB}
\end{equation}
which holds for all three ensembles.
With a little more effort we can evaluate also the variances $\dis{X}\equiv\ave{X^2}-\ave{X}^2$ of these coefficients.
Assuming $d\!\gg\!1$ and taking $\sigma^2$ from \Eqref{sig}, we obtain 
\begin{equation}
\dis{M_V}\approx\left\{\begin{array}{ll}
D_E(0)/d&{\rm for\ }\hat{V}^{\rm diag}\,,\\
2D_E(0)/d^2&{\rm for\ }\hat{V}^{\rm full}\,,\\
0&{\rm for\ }\hat{V}^{\rm offd}\,,\\
\end{array}\right.
\label{MVdis}
\end{equation}
\vspace{-5mm}
\begin{equation}
\dis{D_V}\approx\left\{\begin{array}{ll}
\kappa D^2_E(0)/d&{\rm for\ }\hat{V}^{\rm diag}\,,\\
D^2_E(0)&{\rm for\ }\hat{V}^{\rm full}\,,\\
2D^2_E(0)/d^2&{\rm for\ }\hat{V}^{\rm offd}\,,\\
\end{array}\right.
\label{DVdis}
\end{equation}
\vspace{-5mm}
\begin{equation}
\dis{K}\approx\left\{\begin{array}{ll}
4D^2_E(0)/d&{\rm for\ }\hat{V}^{\rm diag}\,,\\
8D^2_E(0)/d^2&{\rm for\ }\hat{V}^{\rm full}\,,\\
0&{\rm for\ }\hat{V}^{\rm offd}\,,\\
\end{array}\right.
\label{Kdis}
\end{equation}
where $\kappa\=2$ for Gaussian and $\kappa\=0.8$ for rectangular distribution of diagonal matrix elements in the ensemble $\hat{V}^{\rm diag}$.
In these formulas we consider only the leading terms in dimension $d$.

Equations \eqref{AvAB}--\eqref{Kdis} have the following implications:
The average slope $M_V$ of the spectrum for all random interaction ensembles has a zero expectation value, and its expected deviation to up or down direction for a single realization of $\hat{V}$ has a typical value $\abs{\delta M_V}\!\propto\!d^{-1/2}$ for $\hat{V}^{\rm diag}$, $\abs{\delta M_V}\!\propto\!d^{-1}$ for $\hat{V}^{\rm full}$, and $\abs{\delta M_V}\=0$ for $\hat{V}^{\rm offd}$.
The point $\lambda_0$, where the spectrum becomes maximally compressed, is also centered at zero expectation value and its typical deviation to either side for a single random matrix is $\abs{\delta\lambda_0}\!\propto\!d^{-1/2}$ for $\hat{V}^{\rm diag}$, $\abs{\delta\lambda_0}\!\propto\!d^{-1}$ for $\hat{V}^{\rm full}$, and $\abs{\delta\lambda_0}\=0$ for $\hat{V}^{\rm offd}$.
A halfwidth of the minimum of the quadratic spread dependence, i.e., a value $\lambda_1$ such that $D(\lambda_0\!\pm\!\lambda_1)\=2D(\lambda_0)$, is centered at $\lambda_1\approx 1$ irrespective of dimension and class of perturbation, but a typical fluctuation in a single realization  behaves as $|\delta\lambda_1|\!\propto\!d^{-1/2}$ for $\hat{V}^{\rm diag}$, $|\delta\lambda_1|\!\propto\!d^0$ for $\hat{V}^{\rm full}$ and $|\delta\lambda_1|\!\propto\!d^{-1}$ for $\hat{V}^{\rm offd}$.

\section{Level crossing formula for diagonal Hamiltonians}
\label{SeApD}

Here we derive formula \eqref{Decro} for the distribution of level crossings for $\hat{V}\=\hat{V}^{\rm diag}$.
If both $\hat{H}(0)$ and $\hat{V}$ are diagonal matrices, the eigenvalues of $\hat{H}(\lambda)$ are linear functions $E_n(\lambda)\=E_n(0)\+\lambda V_{nn}^{\rm diag}$.
Consider first just a single pair of levels with unperturbed energies $E$ and $E'\=E\+\Delta$ (where $\Delta\>0$) and with random slopes $V$ and $V'$ described by probability densities $P(V)$ and $P(V')$.
The probability to find the crossing of both levels within an interval $\lambda\in[0,\Lambda]$ is trivially determined by:
\begin{equation}
{\cal N}(\Lambda)\=
\int\limits_{-\infty}^{+\infty}\!\!dV\!\!\!
\int\limits_{-\infty}^{V\-\frac{\Delta}{\Lambda}}\!\!\!dV' 
\ P(V)P(V').
\label{A2lev}
\end{equation}
We see that $\lim_{\Lambda\to\infty}{\cal N}(\Lambda)\=1/2$, which expresses the 50\,\% chance to find the crossing at $\lambda\>0$ or $\lambda\<0$. 

This derivation can be easily extended to a general dimension $d$.
Let $\rho(E)=\sum_{n=1}^d\delta(E\-E_n(0))$ is the level density of $\hat{H}(0)$.
The expected number of crossings contained between $\lambda\=0$ and $\Lambda$ is:
\begin{equation}
{\cal N}(\Lambda)\=\!\!
\int\limits_{-\infty}^{+\infty}\!\!dE\!\! 
\int\limits_{0}^{+\infty}\!\!d\Delta\!\!
\int\limits_{-\infty}^{+\infty}\!\!dV\!\!\!\!
\int\limits_{-\infty}^{V\-\frac{\Delta}{\Lambda}}\!\!\!\!dV' 
\rho(E)\rho(E\+\Delta)P(V)P(V'),
\label{Adlev1}
\end{equation}
so the density of crossings ${\cal D}(\lambda)\=\left.\frac{d}{d\Lambda}{\cal N}(\Lambda)\right|_{\Lambda=\lambda}$ reads as:
\begin{equation}
{\cal D}(\lambda)\=\!\!
\int\limits_{-\infty}^{+\infty}\!\!dE\!\! 
\int\limits_{0}^{+\infty}\!\!d\Delta\!\!
\int\limits_{-\infty}^{+\infty}\!\!dV\
\frac{\Delta}{\lambda^2}\,\rho(E)\rho(E\+\Delta)P(V)P\!\left(\!V\!\-\!\frac{\Delta}{\lambda}\!\right),
\label{Adlev2}
\end{equation}
or after the insertion of the discrete expression for $\rho(E)$
\begin{equation}
{\cal D}(\lambda)\=\sum_{n=1}^{d}\sum_{n'=n+1}^{d}
\underbrace{\frac{\Delta_{nn'}(0)}{\lambda^2}\int\limits_{-\infty}^{+\infty}\!\!dV\,P(V)P\!\left(\!V\!\-\!\frac{\Delta_{nn'}(0)}{\lambda}\!\right)}_{{\cal D}_{nn'}(\lambda)}
\label{Adlev3}
\end{equation}
with $\Delta_{nn'}(0)\=E_{n'}(0)\-E_n(0)$.
Introducing a distribution $p(v)\=V_0P(vV_0)$ of dimensionless slopes $v\=V/V_0$, where $V_0$ is a characteristic scale of matrix elements $\hat{V}^{\rm diag}_{nn}$, the summed terms  in \Eqref{Adlev3} are transformed to:
\begin{equation}
{\cal D}_{nn'}(\lambda)=\underbrace{\frac{2V_0}{\Delta_{nn'}(0)}}_{\alpha_{nn'}}\,\underbrace{\frac{2}{\tilde{\lambda}^2}\int\limits_{-\infty}^{+\infty}dv\ p(v)\ p\!\left(v\-\frac{2}{\tilde{\lambda}}\right)}_{F(\tilde{\lambda})}
\label{Adlev4}
\end{equation}
with $\tilde{\lambda}\=\alpha_{nn'}\lambda$.
Thus the contribution to \Eqref{Adlev3} from each level pair is given by a scaled expression ${\cal D}_{nn'}(\lambda)\=\alpha_{nn'}F(\alpha_{nn'}\lambda)$, where $F(\tilde{\lambda})$ is a universal dependence derived from the distribution $p(v)$ and $\alpha_{nn'}$ a scaling factor inversely proportional to the spacing $\Delta_{nn'}(0)$.

Finally, as the integration of ${\cal D}(\lambda)$ over $\lambda\in[0,\infty)$ gives a half of the total number ${\cal I}\=d(d\-1)/2$ of all crossings, we define a $d$-independent distribution of crossings ${\cal P}(\lambda)\=2{\cal D}(\lambda)/{\cal I}$ normalized to a unit integral over the positive axis (the distribution of $\lambda\<0$ crossings is mirror symmetric).
We therefore arrive at \Eqref{Decro}.
The validity of this formula was tested numerically.
The particular forms \eqref{Fgr} of the function $F$ can be easily derived from \Eqref{Adlev4} by inserting the rectangular and normal distributions $p(v)$ with $V_0\=\sqrt{3}\sigma$.



\bibliographystyle{apspre}

\end{document}